\documentclass[%
 reprint,
superscriptaddress,
amsmath,amssymb,
aps,
prb,
longbibliography,nofootinbib
]{revtex4-1}

\usepackage{graphicx}
\usepackage{dcolumn}
\usepackage{bm,bbm,mathrsfs}
\usepackage{xcolor}
\usepackage[colorlinks=true,linkcolor=blue,citecolor=blue]{hyperref}
\newcommand{\ket}[1]{|#1\rangle}
\newcommand{\bra}[1]{\langle#1|}

\newcommand{\inp}[2]{\langle#1|#2\rangle}

\newcolumntype{L}[1]{>{\raggedright\arraybackslash}p{#1}}
\newcolumntype{C}[1]{>{\centering\arraybackslash}p{#1}}
\newcolumntype{R}[1]{>{\raggedleft\arraybackslash}p{#1}}

			\newcommand{\e}[1]{\begin{align}{#1}\end{align}}

		\newcommand{\q}[1]{Eq.\ (\ref{#1})}

		\newcommand{\ri}{\rightarrow}

		\newcommand{\Z}{\mathbb{Z}}

\newcommand{\bn}{\boldsymbol{n}}

        \definecolor{AAcolor}{rgb}{0.7,0.1,0.4}

\newcommand{\bpm}{\begin{pmatrix}}
\newcommand{\epm}{\end{pmatrix}}

\newcommand{\bal}{\begin{align}}

\begin{document}

\preprint{APS/123-QED}

\title{$\mathbbm{Z}_{2}$ Spin Hopf Insulator: Helical Hinge States and Returning Thouless Pump}

 \author{Penghao Zhu} \affiliation{Department of Physics and Institute for Condensed Matter Theory, University of Illinois at Urbana-Champaign, Urbana, Illinois 61801, USA}
  \author{A. Alexandradinata}
  \affiliation{Physics Department, University of California Santa Cruz, Santa Cruz, CA 95064, USA}
 \author{Taylor L. Hughes} \affiliation{Department of Physics and Institute for Condensed Matter Theory, University of Illinois at Urbana-Champaign, Urbana, Illinois 61801, USA}

\date{\today}

\begin{abstract}
We introduce a time-reversal-symmetric analog of the Hopf insulator that we call a spin Hopf insulator. The spin Hopf insulator harbors nontrivial Kane-Mele $\Z_2$ invariants on its surfaces, and is the first example of a nonmagnetic delicate topological insulator with spin-orbit coupling. We show that the Kane-Mele $\Z_2$ topology on the surface is generically unstable, but can be stabilized by the addition of a composition of the particle hole and spatial inversion symmetry. Such a symmetry not only protects the surface $\Z_2$ invariant, but also protects gapless helical hinge states on the spin Hopf insulator. Furthermore, we show that in the presence of four-fold rotational symmetry, the spin Hopf insulator exhibits a returning Thouless pump, as well as surface states on sharp boundary terminations.
\end{abstract}

\maketitle
\section{Introduction}
The Hopf insulator is a three-dimensional bulk insulator that has delicate topology, i.e.,  the topology is unstable to the addition of
 trivial valence or \emph{conduction} bands \cite{PhysRevLett.101.186805,aleks2020multicellularity}. The bulk topology of the Hopf insulator can be diagnosed by the \emph{Hopf invariant}, which is a Brillouin Zone(BZ) integral of the Abelian Chern-Simons three-form \cite{PhysRevLett.101.186805}:
\begin{equation}
\label{eq:hopfnumber}
\chi=-\frac{1}{4\pi^{2}}\int_{BZ}d^{3}k\bm{A}(\bm{k})\cdot\bm{\mathcal{F}}(\bm{k}),
\end{equation}
where $\bm{A}(\bm{k})=i\inp{u}{\nabla_{\bm{k}}u}$ is the Berry connection, $\bm{\mathcal{F}}(\bm{k})=\nabla \times \bm{A}(\bm{k})$ is the Berry curvature, and $\vert u\rangle$ are the occupied Bloch functions. 
For Eq. \ref{eq:hopfnumber} to be gauge invariant, the Chern number of any two-dimensional momentum slice in the three-dimensional BZ must vanish. 

\begin{figure}[h]
\centering
\includegraphics[width=1\columnwidth]{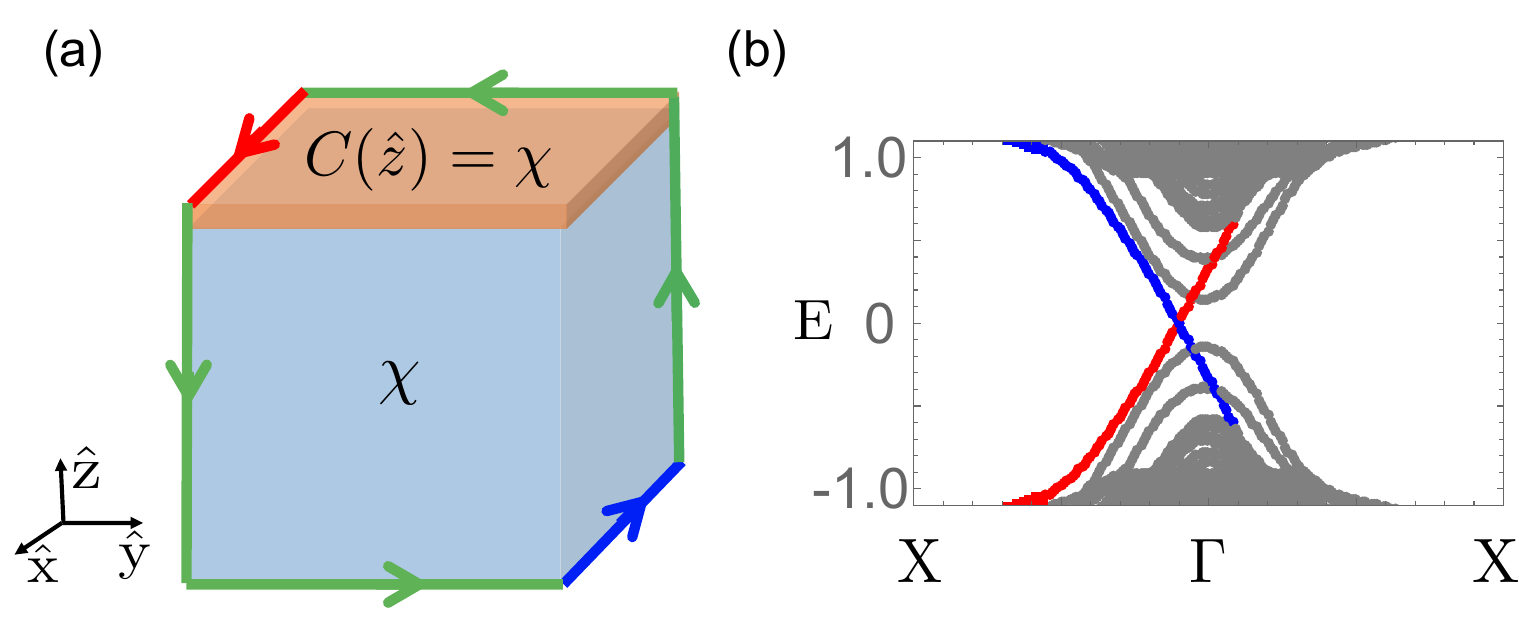}
\caption{(a) For a Hopf-insulating crystallite, the Hopf invariant $\chi$ characterizes the bulk, and the surface Chern number $C(\hat{\bn})$ characterizes each facet and equals the Hopf invariant. If the surface Chern bands are fully occupied for the three forward-facing facets, and if the surface Chern bands are fully unoccupied for the remaining backward-facing facets, then there will be chiral hinge states on the hinges. (b) shows the energy spectrum
for the Hopf insulator in the same geometric configuration as (a) which has open boundaries in $y$, $z$ and periodic
in $x$. The red-colored (blue-colored) in-gap chiral modes is
(are) localized on the red-colored (blue-colored) hinge(s) in
(a). Green-colored hinges also host chiral hinge modes which are,
however, not shown in this energy-momentum plot.  }
\label{fig:hopfillu}
\end{figure}

It has been recently demonstrated that the Hopf insulator has an unusual bulk-boundary correspondence  \cite{alexandradinata2019actually} that relates the bulk Hopf invariant $\chi$ to the Chern number on surface facets having normal vector $\hat{\bm{n}}$ via:
\begin{equation}
\label{eq: bulkboundary}
\begin{aligned}
&C(\hat{\bm{n}})\equiv\int_{rBZ}d^2 k \operatorname{Tr}\left[\bm{\mathcal{F}}_{\hat{\bm{n}}}(\bm{k})\cdot\hat{\bm{n}}\right]=\chi,
\end{aligned}
\end{equation}
where  $\bm{\mathcal{F}}_{\hat{\bm{n}}}(\bm{k})$ is the Berry curvature of boundary-localized bands,
``$\operatorname{Tr}$" is over  \emph{all} boundary-localized bands \footnote{We will explicitly describe what we mean by boundary-localized bands below.} (including both occupied and unoccupied bands), and $rBZ$ indicates the reduced Brillouin zone (rBZ) of the surface facet. We dub $C(\hat{\bm{n}})$ the \emph{surface Chern number}, and it is equal to the Hopf invariant $\chi$ for any orientation of $\hat{\bm{n}}$. Unlike conventional topological insulators, the Hopf bulk-boundary correspondence indicates that the sum over \emph{all} boundary-localized bands has non-trivial first Chern number, instead of just the \emph{occupied} boundary-localized bands. Furthermore, we recently showed that the correspondence between the surface Chern number and the Hopf invariant also implies the existence of chiral hinge states if a spectral gap exists for excitations of surface states \cite{zhu2020quantized}. Thus, the topology diagnosed by the Hopf invariant potentially has a higher order character \cite{benalcazar2018prb,schindler2018higher}. An illustrative picture of the Hopf invariant and surface Chern number correspondence, and the hinge states of a Hopf insulator is shown in Fig.~\ref{fig:hopfillu}.

In addition to the novel bulk-boundary correspondence, rotation symmetric Hopf insulators have been shown to generate a returning Thouless pump (RTP) \cite{alexandradinata2019actually,aleks2020multicellularity,nelson2021delicate}.  A RTP is a newly discovered symmetry-protected property of some delicate topological insulators. In general, an RTP describes charge adiabatically pumped
by a multiple of the lattice period during the first half of the pump, and during the second half of the pump the charge returns to its original position  \cite{alexandradinata2019actually, aleks2020multicellularity,nelson2021delicate}.  Interestingly, the phase transition between a delicate TI with a RTP and a trivial insulator is through a novel topological band degeneracy called a Berry-dipole band degeneracy \cite{aleks2020multicellularity,nelson2021delicate}.

Inspired by these properties of the Hopf insulator, we aim to find a spin-1/2, time-reversal (TR) symmetric analog of the Hopf insulator, which will be the first known example of a delicate topological insulator in Altland-Zirnbauer class AII -- the symmetry class of spin-orbit-coupled time-reversal invariant materials. 
Since non-magnetic materials are more abundant than magnetic materials in nature, the TR symmetric analogs of the Hopf insulator could be easier to realize
in experiments. For these reasons, in this paper, we explore a variety of  phenomena of the TR symmetric analog of the Hopf insulator. For convenience, we will denote the TR symmetric analog of the Hopf insulator as a spin Hopf insulator. In analogy to previous work on Hopf
insulators 
we expect the spin Hopf insulator to have a nontrivial $\Z_{2}$ Kane-Mele invariant \cite{kane2005quantum,PhysRevLett.95.146802,PhysRevLett.96.106802,PhysRevB.79.195321,PhysRevLett.98.106803,qi2008topological,kitaev2009periodic,RevModPhys.88.035005} on the surface, and to exhibit a TR symmetric RTP protected by rotation symmetries.

The remainder of the article is organized as follows. In Sec.~\ref{sec:model} we introduce a model for the spin Hopf insulator and its surface $\Z_2$-invariant. Then in Sec.~\ref{sec:z2pump} we show that through an adiabatic, symmetry-preserving process one can generically eliminate the surface $\Z_2$-invariant by pumping the surface Kane-Mele $\Z_2$-invariant across the bulk via a gapless, Bloch-Wannier band transition. However, in Sec.~\ref{sec:stablesym} we show that with the addition of a particle hole symmetry combined with spatial inversion symmetry the surface $\Z_2$ invariant can be stabilized. Indeed, we go on to show that this symmetry and the nontrivial surface $\Z_{2}$ invariant  can lead to stable helical hinge states. Finally, in Sec.~\ref{sec:rtp} we discuss the RTP protected by four-fold rotation symmetry, and the special band-degeneracy that appears at the topological phase transition point between the rotation invariant spin Hopf insulator and a trivial insulator.

\section{Model Hamiltonian of spin Hopf insulator and surface $\Z_2$-invariant}\label{sec:model} In order to construct a model Hamiltonian of the spin Hopf insulator let us first briefly review the model Hamiltonian of the Hopf insulator constructed by Moore, Ran and Wen (MRW) \cite{PhysRevLett.101.186805}. The MRW Bloch Hamiltonian is given by:
\begin{equation}
\label{eq: MRW}
\begin{aligned}
&z=(z_1+iz_2,z_3+iz_4)^{T},
\\
&\mathbf{d}=z^{\dag}\bm{\sigma}z, \ \bm{\sigma}=(\sigma_{x},\sigma_{y},\sigma_{z}),
\\
&H_{MRW}(\mathbf{k})=\mathbf{d}\cdot\bm{\sigma},
\end{aligned}
\end{equation}
where 
\begin{equation}
\label{eq:z1234}
\begin{aligned}
&z_{1}=\sin k_{x}, z_{2}=\sin k_{y}, z_{3}=\sin k_{z},
\\
&z_{4}=u-\cos k_{x}-\cos k_{y}-\cos k_{z}.
\end{aligned}
\end{equation}
If we take $1<u<3$, then the above model is a Hopf insulator with Hopf invariant $\chi=1$. Note that in the MRW model, any two-dimensional slice of the BZ has trivial Chern number, which indeed is required for $\chi$ to be gauge invariant under gauge transformations of the Bloch wavefunctions.

\begin{figure}[h]
\centering
\includegraphics[width=1\columnwidth]{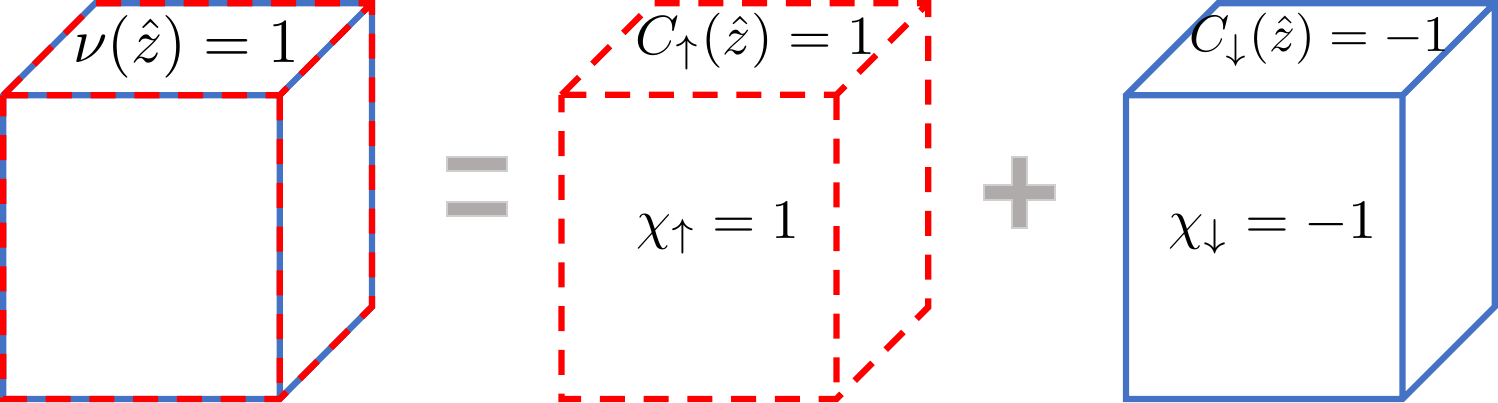}
\caption{Illustration of the construction of spin-Hopf insulators from two copies of Hopf insulators in opposite spin sectors}
\label{fig:spinhopfillu}
\end{figure}

Being an integral of the Chern-Simons three-form [cf.\ \q{eq:hopfnumber}],  $\chi$ transforms like a pseudoscalar under crystallographic spatial transformations and is odd under TR. Thus, TR symmetry is incompatible with a nontrivial Hopf invariant $\chi$. To make progress on a spin Hopf insulator, we consider an enlarged Hilbert space that is the tensor product of the original Hilbert space with a spin-half Hilbert space. In this case, TR symmetry interchanges spin with the representation $T=-i\tau_{y}\mathbbm{1} K,$  and squares to minus the identity, where $K$ is the complex conjugate operator and $\tau_{x,y,z}$ are Pauli matrices for spin. To construct a model that is invariant under $T$, we note that the MRW model Hamiltonian $H_{MRW}$ ($1<u<3$) is a Hamiltonian with $\chi=1$, and replacing $z_{2}$ by $-z_{2}$ in  $H_{MRW}$ leads to a new Hamiltonian $H_{MRW}^{\prime}$ with $\chi=-1$ \cite{nelson2021delicate}. Hence, we can use $H_{MRW}$ for the spin-up sector and  $H_{MRW}^{\prime}$ for the spin-down sector [see Fig.~\ref{fig:spinhopfillu}] to form the direct sum 
\begin{equation}
\label{eq:spinhopf}
\begin{aligned}
H_{sh}(\mathbf{k})&=H_{MRW}\oplus H_{MRW}^{\prime}
\\
&=2 z_{1}z_{3} \mathbbm{1} \sigma_{x}+2 z_{2} z_{4} \tau_{z} \sigma_{x}-2 z_{2}z_{3} \tau_{z} \sigma_{y}
\\
&+2 z_{1} z_{4} \mathbbm{1} \sigma_{y}+\left(z_{1}^{2}+z_{2}^{2}-z_{3}^{2}-z_{4}^{2}\right) \mathbbm{1}\sigma_{z},
\end{aligned}
\end{equation}
where $z_{i}$ for $i=1,2,3,4$ is defined in Eq.~\eqref{eq:z1234}, and $\sigma_{x,y,z}$ ($\tau_{x,y,z}$) are Pauli matrices for orbital (spin), and we have left the tensor product between Pauli matrices implicit. We see that $H_{sh}(\mathbf{k})$ satisfies $TH_{sh}(\mathbf{k})T^{-1}=H_{sh}(-\mathbf{k})$ as expected.

\begin{figure*}[t]
\centering
\includegraphics[width=2\columnwidth]{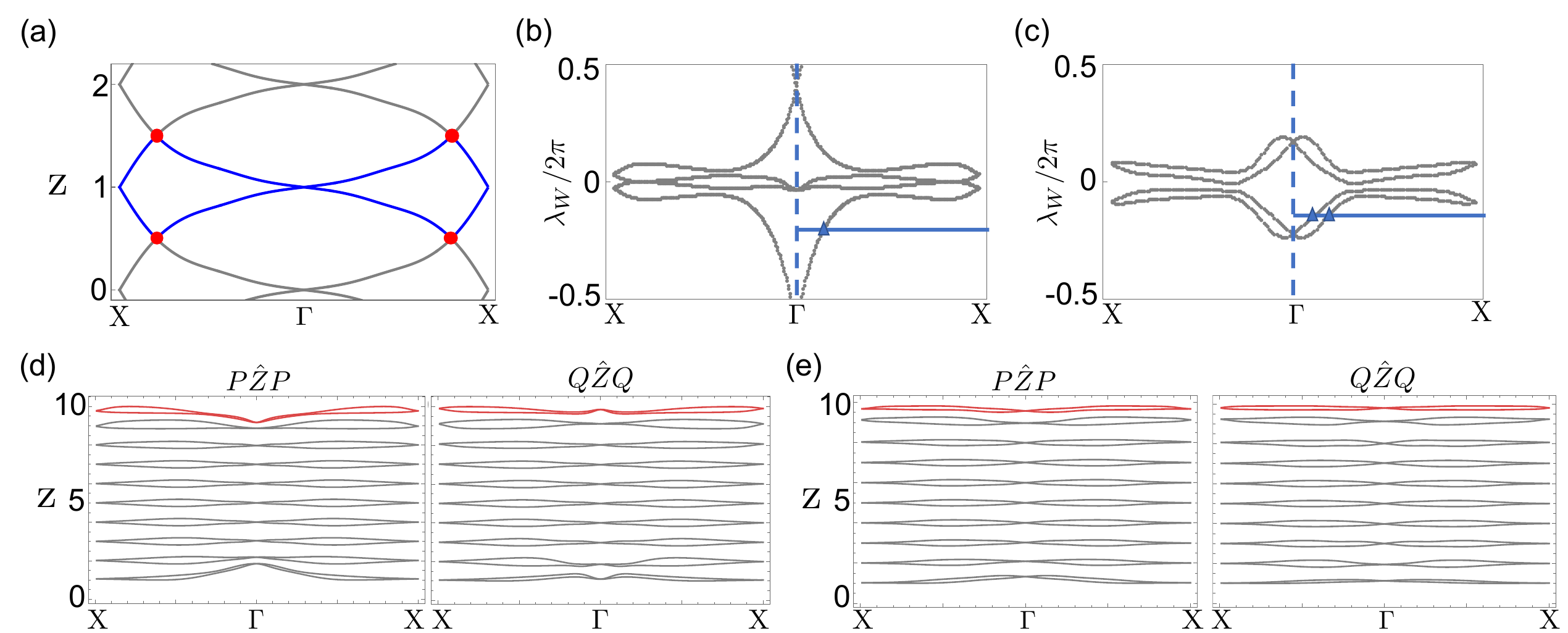}
\caption{(a) A section of the  bulk Bloch-Wannier bands for the spin-Hopf model at the $\Z_{2}$ phase transition point. The Kramers pair of Bloch-Wannier bands indicated by the blue color has touched  its top and bottom neighbors at Dirac-like points (indicated by red dots) related by TR symmetry on the $\rm{X}-\Gamma-\rm{X}$ line in the BZ. (b) and (c) show the spectra of Wilson loops built from projectors of the occupied and unoccupied surface Bloch-Wannier bands (i.e., the layer closest to $z=10$). The Wilson loops are parallel transported along the $k_{y}$-direction and then plotted vs $k_x.$ (b) and (c) show the results  \emph{before} and \emph{after} the $\Z_{2}$ phase transition point. The bands that contribute to the Wilson loop projectors are highlighted in red in (d) and (e) for the calculations done before and after the transition respectively. In (b) and (c) we show the intersections between the Wilson loop spectra and an arbitrary  $\Gamma X$ line (colored blue), which we have indicated by blue triangles. The parity of the number of intersections tells us the $\Z_{2}$ Kane-Mele invariant. We find an odd number in (b) and an even number in (c) indicating their respective non-triviality and triviality. Note that we have used 10 unit cells in the $z$-direction for these calculations.}
\label{fig:pump_z2}
\end{figure*}

Since the opposite Hopf invariants of each spin sector generate opposite surface Chern numbers [cf. Eq.\eqref{eq: bulkboundary}],  the spin Hopf insulator inherits a nontrivial  \textit{surface $\mathbb{Z}_{2}$ invariant} --  defined as the $\mathbb{Z}_{2}$ Kane-Mele invariant \cite{PhysRevLett.95.146802} of all boundary-localized Bloch-Wannier
bands,  as illustrated in Fig.~\ref{fig:spinhopfillu}. To better understand this point, let us clarify what we mean by Bloch-Wannier
bands here. Without loss of generality, we consider a slab geometry with $z$-direction open and $x,y$-directions periodic. We assume that a gap exists in the energy spectrum over the rBZ of the surface facet, and separates two orthogonal subspaces with projection operators $P$ (occupied subspace) and $Q$ (unoccupied subspace). By diagonalizing the projected position operators, $P\hat{Z}P$ and $Q\hat{Z}Q$, where $\hat{Z}$ is the position operator in the $z$-direction, we can obtain the eigenstates which form a set of Bloch-Wannier bands \cite{PhysRevB.89.115102,PhysRevB.83.035108}. These bands are extended (in the $xy$-plane) as Bloch functions with crystal wavevector $\bm{k}_{\perp}=(k_{x},k_{y})$, but exponentially localized as Wannier functions in the $z$-direction. The eigenvalues of $P\hat{Z}P$ and $Q\hat{Z}Q$ represent the Wannier center location in the $z$-direction (see  Fig.~\ref{fig:pump_z2} (d) and (e) for two examples). 

Now let us consider the topological properties of the surface Bloch-Wannier bands. For simplicity, we take a semi-infinite slab with only one surface facet, which has normal vector $+\hat{z}$, and we label the layers along the $z$-direction by $z=1,2,3\ldots$ with $z=1$ being closest to the surface, $z=2$ the next closest, and so on. In each layer, there are two Bloch Wannier bands related by the TR symmetry (see Fig.~\ref{fig:pump_z2} (d) and (e)). Far away from the surface the Bloch-Wannier bands are indistinguishable (up to exponentially small corrections) from Bloch-Wannier bands obtained with periodic boundary conditions in all three spatial coordinates, since the projectors $P, Q$ decay exponentially in the coordinate representation \cite{PhysRev.135.A685}. Hence we dub these bands \emph{bulk} Bloch-Wannier bands, and all other Bloch-Wannier bands are \emph{boundary-localized} bands. We will denote the  Kane-Mele $\Z_2$ invariant of the pair of Bloch-Wannier bands in the \emph{occupied} subspace and localized in the layer labelled by $z$ as $\nu_{v}(+\hat{z},z)$, and define the surface valence $\Z_{2}$ invariant as
\begin{equation}
\label{eq: surfz2v}
\nu_{v}(+\hat{z})=\sum_{z=1}^{bulk}\nu_{v}(+\hat{z},z),
\end{equation} 
where $\sum_{z=1}^{bulk}$ means that the sum extends deep enough into the slab to the region containing bulk Bloch-Wannier bands. In other word, ``deep enough" means that it is much deeper than the localization length of the boundary modes that is determined by the inverse of bulk energy gap. The sum in Eq.~\eqref{eq: surfz2v} is uniquely defined because bulk bands have trivial $\Z_{2}$ Kane-Mele invariant, which is inherited from the vanishing Chern number of the bulk bands of the Hopf insulator in each spin sector. Similarly, we can define  $\nu_{c}(+\hat{z},z)$ and the surface conduction $\Z_2$ invariant $\nu_{c}(+\hat{z})$ for the unoccupied subspace.

All the definitions discussed above can be extended to any insulating surface facets by replacing $+\hat{z}$ with a generic $\hat{\bm{n}}$ representing the normal vector of the surface facet of interest.  Then, as defined above, the \emph{surface $\Z_{2}$ invariant} for a surface facet with normal vector $\hat{\bm{n}}$ is the  $\Z_{2}$ Kane-Mele invariant of \emph{all} boundary-localized bands, i.e.,
\begin{equation}
\label{eq: surfz2}
\nu({\hat{\bm{n}}})=\nu_{v}({\hat{\bm{n}}})+\nu_{c}({\hat{\bm{n}}}).
\end{equation}
Note that $\nu_{\hat{\mathbf{n}}}$ can also be calculated when there is no surface spectral gap using the method discussed in Ref.~\onlinecite{Trifunovic2020}

\section{Trivialization from Pump of $\Z_2$ Kane-Mele invariant}\label{sec:z2pump} While the explicit model we wrote in Eq.~\eqref{eq:spinhopf} has a nontrivial surface $\Z_2$ invariant inherited from the surface Chern numbers of the Hopf insulators for each spin sector, we must be careful to understand the stability of this nontrivial surface $\Z_2$ invariant under gap and symmetry-preserving deformations.
Indeed, we find that with no additional symmetries other than TR, the spin Hopf insulator can be continuously connected to a trivial band insulator while preserving the bulk \textit{energy} gap. We provide an explicit symmetry and gap-preserving interpolation from the model Hamiltonian in \q{eq:spinhopf} to a trivial insulator in the Appendix~\ref{app:adiabaticpath}. This is consistent with an earlier homotopic classification in Ref.~\onlinecite{kennedy2014homotopy} which asserts there are no delicate topological insulators in Altland-Zirnbauer class AII (in the first order topology sense).

Interestingly, even though the bulk \emph{energy} gap is maintained during our interpolation, we expect that the spectral gap of the bulk \textit{Bloch-Wannier} bands \emph{must} close at some point to effect a change in the surface $\Z_{2}$ invariant. Indeed, this gap closing realizes a real-space pump of the $\Z_{2}$ Kane-Mele invariant, i.e., one quantum of the $\Z_{2}$ Kane-Mele invariant leaves the surface and propagates through the bulk. If this propagation occurs in the occupied (resp.\ unoccupied) bulk states, it is the bulk spectral gap of $P\hat{Z}P$ (resp.\ $Q\hat{Z}Q$) that closes. For a finite slab of the spin Hopf insulator, this indicates that the $\Z_{2}$ Kane-Mele invariant will be pumped from each surface and annihilate in the bulk, leading to a net cancellation of the surface $\Z_2$ invariant on both surfaces -- this is the TR-symmetric analog of the Berry-curvature teleportation that occurs at the critical transition between trivial and Hopf insulators \cite{alexandradinata2019actually}.

With this phenomenology in mind, let us illustrate this pump of the $\Z_2$ Kane-Mele invariant on a slab geometry that is open in the $z$-direction and periodic in the $x,y$-directions. In our interpolation we couple the two spin sectors of \q{eq:spinhopf} and modify the surface Hamiltonian to create an energy gap between the occupied and unoccupied subspaces. This surface energy gap will remain throughout the part of our interpolation in which the $\Z_2$ pump occurs. For our interpolated Hamiltonian [see Appendix~\ref{app:adiabaticpath}], the topologically nontrivial Bloch-Wannier bands on the top and bottom surface that participate in the $\Z_2$ pump lie in the occupied subspace. Therefore, it is sufficient to focus our attention on $P\hat{Z}P$ alone. At the critical point of the interpolation we plot the spectrum of $P\hat{Z}P$ in Fig.~\ref{fig:pump_z2} (a). As we can see, each bulk Bloch-Wannier band in the occupied space touches its top and bottom neighbors at two TR related points on the $\rm{X}-\Gamma-\rm{X}$ high symmetry line (thus each forms four Dirac-like points  as indicated by red dots in Fig.~\ref{fig:pump_z2} (a)).

The transition where the $\Z_2$ pump occurs is analogous to the phase transition of the 2D quantum spin Hall phase \cite{Murakami2007phase,Shuichi_Murakami_2007}, which can happen through a pair of TR related, Dirac-like gap closing points at general momenta $\mathbf{k}$ and $-\mathbf{k}$. In another word, the critical point where the $\Z_2$ pump occurs is a $\mathbbm{Z}_{2}$ phase transition point for the two-dimensional bulk Bloch-Wannier bands.  Because of the translation symmetry in the bulk, each bulk Bloch-Wannier band (indicated by blue in Fig.~\ref{fig:pump_z2} (a)) touches with both its top and bottom neighbors simultaneously, and thus the $\mathbbm{Z}_{2}$ phase transition happens twice for a bulk Bloch-Wannier band -- once from the top and once from the bottom. The net effect is that the $\mathbbm{Z}_{2}$ invariant of the bulk Bloch-Wannier bands will not change during this process. However, the boundary-localized Bloch-Wannier bands on both top and bottom surface facets touch with the bulk Bloch-Wannier bands from only one side, i.e., the boundary-localized Bloch-Wannier bands on the top (bottom)  touch with only the bulk Bloch-Wannier bands below (above) it. Thus, there will be a quantum of $\mathbbm{Z}_2$ invariant, $\nu=1$, pumped from one surface to another through the bulk occupied bands during this process and all of the Bloch-Wannier bands will become trivial. After this critical point, the gaps in the bulk Bloch-Wannier bands reopen and the system will be a trivial insulator. We note that breaking translation symmetry does not change our conclusions, since the full $\Z_2$ pump can still occur, but in perhaps a non-uniform fashion where some of the bulk Bloch-Wannier bands become $\Z_2$ non-trivial during an intermediate stage. We also point out that the $\Z_2$ pump occurs in the $P$ subspace but not in the $Q$ subspace. This should be contrasted with previously-studied pumps associated to a second Chern number, where a Berry-curvature quantum is simultaneously pumped in both $P$ and $Q$ subspaces \cite{Taherinejad2015adiabatic,PhysRevB.95.075137}.


To confirm our results we use the fact that the $\Z_{2}$ invariant can be determined from spectrum of the Wilson loop \cite{PhysRevB.74.195312,bernevig2013topological,asboth2016short}.
 Let $p(\mathbf{k})$ be the projector onto a chosen group of bands at $\mathbf{k}$. We can define the Wilson loop operator over this group of bands as
\begin{equation}
\label{eq: WLO}
\widehat{\mathcal{W}}=\prod_{\mathbf{k}}^{\mathscr{C}}p(\mathbf{k}),
\end{equation}
where $\prod^{\mathscr{C}}_{\mathbf{k}}$ is a path-ordered product over the loop $\mathscr{C}$ in momentum space, and $p_{v}(\bm{k})=\sum_{j=1}^{n_{occ}}\ket{u_{j}(\mathbf{k})}\bra{u_{j}\mathbf{k})}$ with $\ket{u_{j}(\mathbf{k})}$ to be the $j$-th eigenstate of the Bloch Hamiltonian and $n_{occ}$ to be the number of occupied bands. The exponents of all unit modulus complex eigenvalues of the Wilson loop operator are defined to be the spectrum of the Wilson loop. Hereafter, we denote the spectrum of the Wilson loop to be $\lambda_{W}$. To derive the Kane-Mele $\Z_2$ invariant of a given group of bands, we consider $\mathscr{C}$ to be a non-contractible loop along the $k_{y}$-direction, e.g., from $k_{y}=0$ to $k_{y}=2\pi$ at fixed values of $k_{x}$. To determine the invariant we can use the parity of the number of intersections between a reference line with fixed lambda but with $k_{x}$ varied from $0$ to $\pi$ and the spectrum of the Wilson loop: if the number of intersections is odd (even), then $\Z_{2}$ invariant is one (zero) and thus nontrivial (trivial). Note that the reference line can be arbitrarily chosen.

In Fig.~\ref{fig:pump_z2} (b) and (c), we show numerical plots of the Wilson loop spectra along $k_{y}$ (parallel transport along $k_{y}$-direction) for a projector that includes both the occupied and unoccupied Bloch-Wannier bands localized at the top surface (i.e., the layer closest to $z=10$) before and after the phase transition point of the bulk Bloch-Wannier bands respectively. Specifically, before (after) the phase transition point, the parallel transport is performed on the surface-localized occupied and unoccupied Bloch-Wannier bands indicated by the red colored states in Fig.~\ref{fig:pump_z2} (d) (Fig.~\ref{fig:pump_z2} (e)). As shown in Fig.~\ref{fig:pump_z2} (b), the blue reference line parallel to $\Gamma$-$X$ crosses the Wilson loop spectrum at one point indicated by a blue triangle. Thus the surface $\Z_2$ invariant is non-trivial on the top surface before the critical point. By comparison, in Fig.~\ref{fig:pump_z2} (c), the blue reference line parallel to $\Gamma$-$X$ crosses the Wilson loop spectrum at two points indicated by the two blue triangles. Thus, in this case we have a trivial surface $\Z_2$ invariant for the top surface after the critical point. These results are consistent with our previous 
argument, which 
was based on the
Dirac-like touching
points between 
Bloch-Wannier bands. More details about the evolution of the surface $\mathbbm{Z}_{2}$ invariant are given in the Appendix~\ref{app:evolution}.

While our symmetry-preserving interpolation trivializes our model, we note that the topology can remain nontrivial as long as the Bloch-Wannier band gap remains open in addition to the bulk energy gap. Indeed, if the Bloch-Wannier gap is kept open the $\Z_2$ pump will not occur, and the surface $\Z_{2}$-invariant will be a robust indicator of the nontrivial topology. This is similar to the boundary obstructed topology of some electric multipole insulators \cite{benalcazar2017quantized,benalcazar2017electric,khalaf2021,zhu2021multi}, in the sense that the topology is protected by the bulk energy gap together with bulk Bloch-Wannier gap. In the next subsection we will show that instead of requiring that the spectral gap of the Bloch-Wannier bands is preserved, we can impose a symmetry that can generically prevent the pump and stabilize the surface $\Z_{2}$ invariant. We also demonstrate the existence of stable helical hinge modes in the spin Hopf insulator under this symmetry.

\section{symmetry-protected surface $\Z_2$ topology and helical hinge states}\label{sec:stablesym}
\subsection{Particle-hole inversion symmetry}
In this section we will show that the surface $\Z_{2}$ invariant of the TR-symmetric spin-Hopf insulator can be stabilized if we impose an additional particle-hole times spatial inversion symmetry in each spin sector: $\mathcal{C}^{\prime}c_{\bm{R},i,s}\mathcal{C}^{\prime-1}=\epsilon_{ij}c_{-\bm{R},j,s}^{\dag}$, where $i,j$ label the orbital degrees of freedom, $s$ labels the spin, $c_{\bm{R},i,s}^{\dag}$ creates an electron at $\bm{R}$, and $\epsilon_{ij}$ is the two-dimensional Levi-Civita tensor. Note that this symmetry is a spin-doubled version of that discussed in Ref.~\onlinecite{zhu2020quantized}, and is related to a symmetry used in Ref.~\onlinecite{Liu2017symmetry} to extend the Hopf insulator to multiple bands. For simplicity we shorten the name of the symmetry to $\mathcal{C}^{\prime}$ symmetry in the following.
The $\mathcal{C}^{\prime}$ symmetry of the second-quantized Hamiltonian constrains the first-quantized Bloch Hamiltonian to obey an anti-symmetry:
\e{\mathbbm{1}\sigma_{y}H^{\star}_{sh}(\bm{k})\mathbbm{1}\sigma_{y}=-H_{sh}(\bm{k}), }
where $^{\star}$ is the complex conjugate.

With this definition, let us demonstrate that the $\mathcal{C}^{\prime}$ symmetry can stabilize the surface $\Z_{2}$ invariant, i.e.,  any continuous deformations that preserve the symmetries and bulk energy gap \emph{cannot} change the surface $\Z_{2}$ invariant. Note that the particle-hole symmetry is broken by the adiabatic interpolation (see the Appendix~\ref{app:adiabaticpath}) from the model Hamiltonian in \q{eq:spinhopf} to a trivial Hamiltonian, owing to a Hamiltonian  term $tz_{1}\tau_{y}\sigma_{x}$ that is introduced in the interpolation. We note that the explicit interpolation we used in the previous section violates $\mathcal{C}^{\prime}$-symmetry so it is no longer valid. Instead of proving directly that no such interpolation can exist in the presence of $\mathcal{C}^{\prime}$ symmetry, we will take a different approach below to prove the stability of the spin Hopf phase protected by the addition of $\mathcal{C}^{\prime}$.

Our intuition about the nature of the stability comes from the fact that in a slab that is finite along $z$ direction and thus has top and bottom surface facet, if a $\Z_2$ Kane-Mele quantum is pumped from the bottom facet to the top through the bulk \textit{occupied} Bloch-Wannier bands, the $\mathcal{C}^{\prime}$ symmetry enforces that a $\Z_2$ Kane-Mele quantum  will also be pumped from top to bottom through the bulk \textit{unoccupied} Bloch-Wannier bands, and hence the processes would cancel each other.
Without loss of generality, let us illustrate this idea in a spin-Hopf insulator with $z$-direction open, and both $x,y$-directions periodic. 

The $\mathcal{C}^{\prime}$ symmetry will enforce
\begin{equation}
\label{eq:constraint_spinhopf}
\nu_{v}(+\hat{z})=\nu_{c}(-\hat{z}), \ \nu_{c}(+\hat{z})=\nu_{v}(-\hat{z}),
\end{equation}
where we recall that $\nu_{v}(\hat{\bm{n}})$ ($\nu_{c}(\hat{\bm{n}})$) is defined to be the Kane-Mele $\mathbbm{Z}_2$ invariant of boundary-localized occupied (unoccupied) Bloch Wannier bands, and $\nu_{v}(\hat{\bm{n}})=1$ ($\nu_{c}(\hat{\bm{n}})=1$) indicates nontrivial topology in occupied (unoccupied) subspace. Eq.~\eqref{eq:constraint_spinhopf} can be easily understood from the Bloch-Wannier bands. If we have $\mathcal{C}^{\prime}$ symmetry then $P\hat{Z}P=-\mathcal{C}^{\prime}Q\hat{Z}Q\mathcal{C}^{\prime-1}$. Since particle-hole
symmetry interchanges $P$ with $Q$, and inversion
symmetry interchanges $z$ with $-z$, if $\ket{w_{z}^{v}(k_{x},k_{y})}$ is an eigenstate of $P\hat{Z}P$ with eigenvalue $z(k_{x},k_{y})$, then $\mathcal{C}^{\prime-1}\ket{w_{z}^{v}(k_{x},k_{y})}=\ket{w_{-z}^{c}(k_{x},k_{y})}$ is an eigenstate of $Q\hat{Z}Q$ with eigenvalue $-z(k_{x},k_{y})$. Note that the subscript $z$ and $-z$ indicates the corresponding eigenvalues of the eigenstates.
Since the Bloch-Wannier wave functions at $\pm z$ are related by an antiunitary operation that maps $(k_x,k_y)\ri (k_x,k_y)$, they must have the same Kane-Mele index, meaning $\nu_{v}(\pm \hat{z})=\nu_{c}(\mp \hat{z})$. 
A detailed proof following this symmetry argument can be found in the Appendix~\ref{app: proofofeq31}.

In terms of the quantities $\nu_{v}$ and $\nu_{c}$ discussed above, the total surface $\mathbbm{Z}_{2}$ invariant of a surface facet with the normal vector, e.g., $+\hat{z}$, is $\nu(+\hat{z})=\nu_{v}(+\hat{z})+\nu_{c}(+\hat{z})$. Since the bulk Bloch-Wannier bands have trivial $\Z_{2}$ invariant, the full slab $\Z_{2}$ invariant of the occupied (unoccupied) subspace, $\nu_{v/c}$, is  contributed by only the top and bottom surfaces, i.e., $\nu_{v/c}=\nu_{v/c}(+\hat{z})+\nu_{v/c}(-\hat{z}) \ \mathrm{mod}\ 2$.  Now, we are ready to  prove our target conclusion: any continuous deformations that preserve the symmetries and bulk gap \emph{cannot} change the surface $\Z_{2}$ invariant. We separate our proof into two steps. First, if we preserve both the bulk and surface gaps in the energy spectrum, $\nu_{v/c}$ cannot change.  From  Eq.~\eqref{eq:constraint_spinhopf}, we can derive 
\begin{equation}
\label{eq: nuv_nus}
\begin{aligned}
&\nu_{v/c}=\nu_{v/c}(+\hat{z})+\nu_{v/c}(-\hat{z})
\\
&=\nu_{v/c}(+\hat{z})+\nu_{c/v}(+\hat{z})=\nu(+\hat{z})
\\
&=\nu_{c/v}(-\hat{z})+\nu_{v/c}(-\hat{z})=\nu(-\hat{z}).
\end{aligned}
\end{equation}
Thus, if $\nu_{v/c}$ remains invariant, so do both $\nu(+\hat{z})$ and the $\nu(-\hat{z})$. Hence, $\mathcal{C}^{\prime}$ symmetry along with bulk and surface gaps rule out a change of $\nu(+\hat{z})$ and $\nu(-\hat{z})$ due to a $\Z_{2}$ pump that occurs at the gap closing point in the spectra of $P\hat{Z}P$ and $Q\hat{Z}Q$. Second, we relax the requirement for a surface energy gap. Since the surface $\Z_{2}$ invariant is the $\Z_{2}$ Kane-Mele invariant over the rBZ of \emph{all} boundary-localized Bloch-Wannier bands, it will not change  even if the surface gap in the energy spectrum closes and reopens since such a scenario, at worst, simply exchanges the topology between the occupied and unoccupied subspaces.

In conclusion, even if the surface gap is not preserved, the surface $\mathbbm{Z}_{2}$ invariants $\nu(\hat{z})$ and $\nu(-\hat{z})$, will never change as long as the bulk gap, TR symmetry, and $\mathcal{C}^{\prime}$ symmetry are preserved. In other words, in the spin-Hopf insulator with the particle-hole inversion symmetry and TR symmetry, the surface $\mathbbm{Z}_{2}$ invariant is stable against any adiabatic deformations, and can be a well-defined topological invariant. Note that our proof did not use spin-$U(1)$ symmetry of the model, hence the conclusion remains valid with $U(1)$-breaking spin-orbit couplings. Our proof also indicates that the surface $\Z_2$ topology is stable to the simultaneous addition of momentum-independent Kramers pairs of bands (which would have trivial $\Z_2$ topology everywhere) in the valence and conduction subspaces such that the particle-hole inversion symmetry is preserved. This generalizes the spin-Hopf insulator to multi-band cases. We note that the above proof still works if we change $\hat{z}$ to any other surface normal vector $\hat{\bm{n}}$.

\begin{figure}[h]
\centering
\includegraphics[width=1\columnwidth]{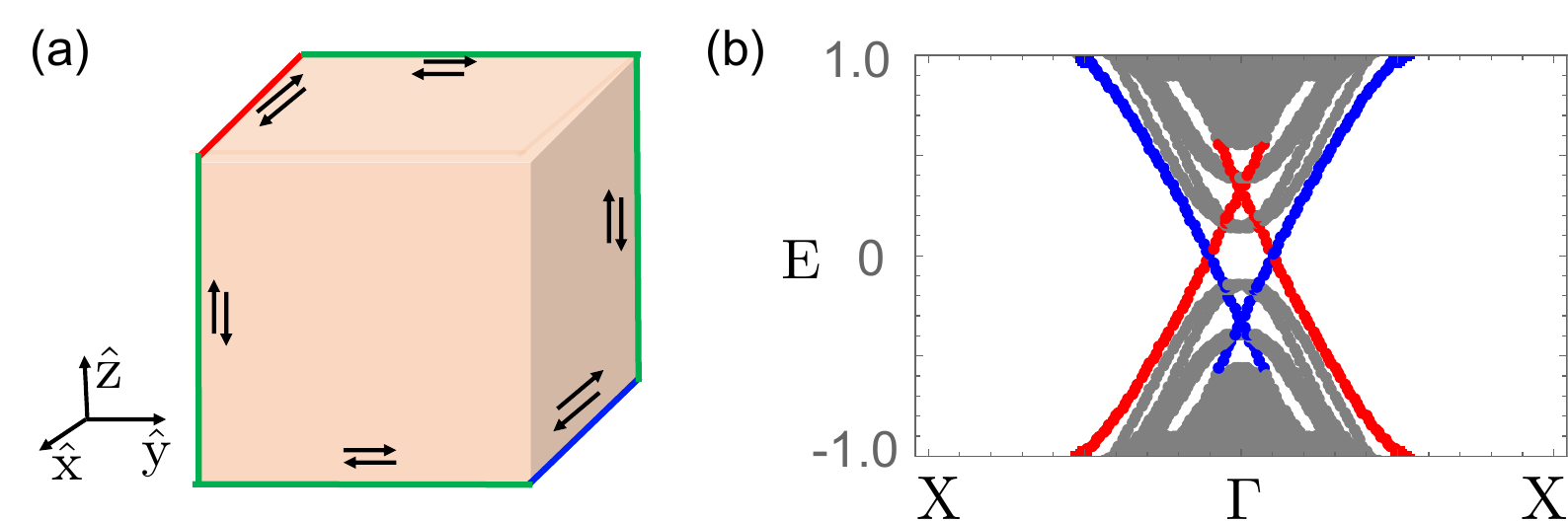}
\caption{(a) An illustration of the helical hinge modes for a TR-symmetric spin-Hopf insulator with nontrivial surface $\mathbbm{Z}_2$ invariant which has $\mathcal{C}^{\prime}$ symmetry. (b) is the energy spectrum for the spin-Hopf insulator in the same geometric configuration as (a) which has open boundaries in $y,z$ and periodic in $x$. The red-colored (blue-colored) in-gap helical modes is (are) localized on the red-colored (blue-colored) hinge(s) in (a). Green-colored hinges also host hinge modes which are, however, not shown in this energy-momentum plot. We used a lattice having 10 unit cells along the $z$-direction and 40 unit cells along the $y$-direction.}
\label{fig:hinge_spinhopf}
\end{figure}

\subsection{Stable helical hinge modes} 
We have now shown that imposing $\mathcal{C}^{\prime}$ symmetry on a TR-symmetric spin Hopf insulator, or just requiring the gap of the bulk Bloch-Wannier bands to be preserved, stabilizes the surface $\Z_2$ invariant. An important implication of the stable surface $\Z_2$ invariant is that there could be gapless helical modes on the hinges where two facets with distinct $\nu_{v}$ meet. It is then natural to ask if these helical modes are stable in the presence of surface perturbations. 

If we only have TR symmetry, even if we require the bulk Bloch Wannier gap to be preserved, we find that the helical hinge modes are not stable. That is,  they can be removed by adding certain surface perturbations  to make all surface facets have the same surface valence $\Z_{2}$ invariant (while preserving the energy gap and the gap of the bulk Bloch-Wannier bands).  This is because the surface valence $\Z_{2}$ invariant can be changed by perturbations on the surface facets that close and reopen the \emph{surface} energy gap. It is only the surface $\Z_{2}$ invariant of both the occupied and unoccupied bands combined that is fixed.

In contrast, the particle-hole inversion symmetry can stabilize gapless helical hinge modes by ensuring the existence of domain walls on the closed two-dimensional surface of our 3D crystallite that is finite in all directions, as long as the surface energy spectrum is gapped. Recall from \q{eq:constraint_spinhopf} that the symmetry enforces $\nu_{v}(\hat{\bm{n}})=\nu_{c}(-\hat{\bm{n}})$ for all $\hat{\bm{n}}$. Then, for spin Hopf insulators with $\nu(\hat{\bm{n}})=1$ for all $\hat{\bm{n}}$ [cf.\ Eq.~\eqref{eq:spinhopf}], we have
\begin{equation}
\nu_{v}(\hat{\bm{n}})-\nu_{v}(-\hat{\bm{n}})=1\ \rm{mod}\ 2.
\end{equation}
Thus, there must be a domain wall between distinct Kane-Mele $\mathbbm{Z}_2$ invariants on the two-dimensional surface of the spin-Hopf insulator, and this domain wall will harbor gapless helical hinge modes. As a representative illustration, Fig.~\ref{fig:hinge_spinhopf} (a) shows a pattern of helical hinge modes for a particular, symmetric surface termination of our model (whose bulk Hamiltonian is given in  Eq.~\eqref{eq:spinhopf}). In Fig.~\ref{fig:hinge_spinhopf} (b) we show the corresponding energy spectrum for our model where periodic boundary conditions have been imposed in the $x$-direction, and open boundaries in both the $y,z$ directions such that there are four surface facets with normals in the $\pm \hat{y}$ and  $\pm \hat{z}$ directions. Perturbations can change the spatial pattern of helical hinge states, but cannot remove them entirely.

\section{Returning Thouless pump protected by a four-fold rotation symmetry}\label{sec:rtp}
\subsection{Theory for the Returning Thouless Pump}
After discussing the nontrivial surface $\Z_2$ topology and helical hinge states in the spin Hopf insulator, in this Section we discuss the crystalline symmetry-protected returning Thouless pump (RTP), of the spin Hopf insulator that is stabilized by a four-fold rotation symmetry.

Let us first review the general theory of the RTP in $n$-fold rotation symmetric insulators \cite{aleks2020multicellularity}. Without loss of generality, let us focus on the $n$-fold rotation symmetry of the lattice along the $z$-direction ($C_{nz}$), and denote the fixed points of the $C_{nz}$ rotation in the $k_{x}-k_{y}$ plane as $\boldsymbol{\Lambda}$, i.e., $C_{nz}\boldsymbol{\Lambda}= \boldsymbol{\Lambda}$ up to a reciprocal lattice vector. At each $\boldsymbol{\Lambda}$, the Bloch Hamiltonian represents a gapped 1D Hamiltonian along the $z$-direction $H_{\boldsymbol{\Lambda}}(k_{z})$. For convenience, we always choose a basis in which $C_{nz}$ is diagonal so that at each $\boldsymbol{\Lambda}$, $H_{\boldsymbol{\Lambda}}(k_{z})$ is always block diagonal, and the blocks are labelled by the eigenvalues of $C_{nz}$. Given that $C_{nz}$ is a unitary operator, the eigenvalues of  $C_{nz}$ generally take the form of $e^{i2\pi l/n}$, where $l$ is the discrete-valued angular momentum. 

Now we can calculate the charge polarization along the $z$-direction contributed by all eigenstates having angular momentum $l$ of the 1D Bloch Hamiltonian at momentum $\boldsymbol{\Lambda}$, and denote it as $\mathscr{P}_{l}(\boldsymbol{\Lambda})$.
Note we are considering the polarization of \emph{all} of the energy bands (both occupied and unoccupied) the Hilbert space spanned by eigenstates of $H_{\boldsymbol{\Lambda}}(k_{z})$ with angular momentum $l$ is independent of $\boldsymbol{\Lambda}$, because it is always the full Hilbert space spanned by basis orbitals with angular momentum $l$ in each unit cell. Therefore, the difference of $\mathscr{P}_{l}$ between a pair of $\boldsymbol{\Lambda}$'s is always quantized to be an integer:
\begin{equation}
\label{eq:pdiff}
\Delta_{12}\mathscr{P}_{l}\equiv\mathscr{P}_{l}(\boldsymbol{\Lambda}_{1})-\mathscr{P}_{l}(\boldsymbol{\Lambda}_{2})=m, \ m\in \Z,
\end{equation}
where $\boldsymbol{\Lambda}_{1}$ and $\boldsymbol{\Lambda}_{2}$ are two different rotation invariant points.
A quantized difference $\Delta_{12}\mathcal{P}$ indicates that if we go from $\boldsymbol{\Lambda}_{1}$ to $\boldsymbol{\Lambda}_{2}$, the polarization along the $z$-direction continuously changes by an integer. Then if we go from $\boldsymbol{\Lambda}_{2}$ to $\boldsymbol{\Lambda}_{1}$ by completing a noncontractible loop across the BZ, the polarization returns back to its original value. Hence, in analogy with the Thouless pump where the polarization changes by an integer over a loop, we say that in this case there is a RTP with value $m$ in sector $l$ along the loop $\boldsymbol{\Lambda}_{1}\to\boldsymbol{\Lambda}_{2}\to\boldsymbol{\Lambda}_{1}$ if $\Delta_{12}\mathscr{P}_{l}=m$.

Let us consider a special case where the eigenstates at rotation invariant points with a given angular momentum $l$ are either all in the occupied subspace or all in the unoccupied subspace (this is the mutually disjoint condition defined in Ref.~\onlinecite{aleks2020multicellularity}), then the polarization of all the occupied bands can be expressed as
\begin{equation}
\label{eq: valenceP}
    \mathscr{P}_{v}(\boldsymbol{\Lambda})=\sum_{l\in l_{v}}\mathscr{P}_{l}(\boldsymbol{\Lambda}),
\end{equation}
where $l_{v}$ indicates the set of all angular momenta in the occupied subspace. Using Eqs. \eqref{eq:pdiff} and \eqref{eq: valenceP}, it is straightforward to see that the difference of $\mathcal{P}_{v}$ between a pair of rotation invariant points is also quantized to be an integer. This quantized difference cannot be changed as long as the bulk energy gap and the $C_{nz}$ symmetry are preserved. Similarly, under these conditions we say there is an RTP with value $m$ in the occupied valence subspace along the loop $\boldsymbol{\Lambda}_{1}\to\boldsymbol{\Lambda}_{2}\to\boldsymbol{\Lambda}_{1}$ if $\Delta_{12}\mathscr{P}_{v}=m.$

\begin{figure}[h]
\centering
\includegraphics[width=1\columnwidth]{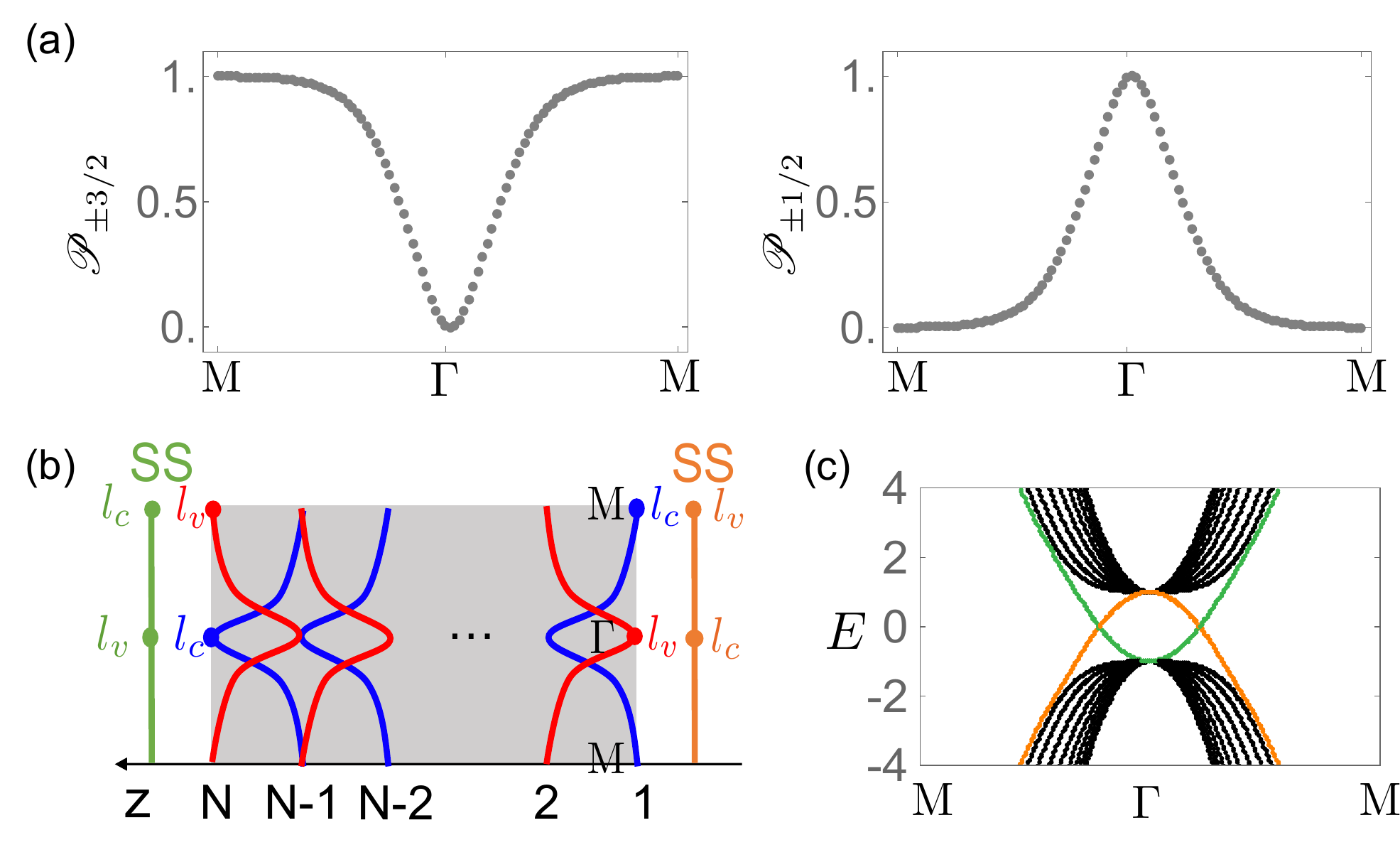}
\caption{(a) An RTP along $M-\Gamma-M$ for valence bands (left) with $l_{v}=\pm 3/2$ and conduction bands (right) with $l_{c}=\pm 1/2$. $\mathscr{P}_{l}$ at generic momenta on the $\Gamma-M-\Gamma$ high symmetry line indicates the polarization along $z$ direction contributed by states on the band with 
$C_{4z}$ eigenvalues $l$ at both $\Gamma$ and $M$. (b) schematically illustration of Bloch-Wannier states protruding from the bulk to the surface (colored red for $l_{v}=\pm 3/2$ and blue for $l_{c}=\pm 1/2$ ). Since the Hilbert space in each layer is fixed there is are compensating surface states (SS) (represented by a green line and an orange line) with the compensating angular momenta at $\Gamma$ and $M$. (c) shows the spectrum of the model with a sharp termination where in-gap SS localized on the top surface, i.e., $z=N$ (highlighted by the green color), and on the bottom surface, i.e., $z=1$ (highlighted by the orange color) are shown.}
\label{fig:RTP}
\end{figure}

Now that we have finished reviewing the general theory of the RTP, we are ready to discuss the RTP for the spin-Hopf insulator. Although we did not emphasize it, Eq.~\eqref{eq:spinhopf} already has a four-fold rotation symmetry along the $z$-direction, for which the operator is $C_{4z}=\exp\left(i\frac{\pi}{2}\tau_{z}(\frac{\sigma_{z}}{2}+\mathbbm{1})\right)$. The two valence (conduction) eigenbands have $C_{4z}$ eigenvalues $\exp(\pm i 3\pi/4)$ ($\exp(\pm i \pi/4)$) respectively, at both $C_{4z}$ symmetric points $\Gamma$ and $M$. This corresponds to $l_{v}=\{3/2,-3/2\}$ ($l_{c}=\{1/2,-1/2\}$). Clearly, the mutually disjoint condition is satisfied because $l_{v}\cap l_{c}=\emptyset$. As shown in Fig.~\ref{fig:RTP} (a), the RTP data in different angular momentum sectors are $\Delta_{\Gamma M}\mathscr{P}_{\pm 3/2}=-1$ and $\Delta_{\Gamma M}\mathscr{P}_{\pm 1/2}=+1$. Thus, the RTP in the valence subspace has a value $\Delta_{\Gamma M}\mathscr{P}_{v}=-2$. Hence, the nontrivial polarization difference at $\Gamma$ and $M$ leads to a RTP, which is robust to perturbations and deformations (e.g., introducing couplings between two spin sectors) as long as we preserve the $C_{4z}$ symmetry and the bulk energy gap. 

\subsection{Implications of the RTP for Gapless surface states on sharp terminations}
Delicate TIs with a nonzero RTP exhibit an unusual bulk-boundary correspondence as discussed in Ref.~\onlinecite{nelson2021delicate}. As an example of this, in this subsection, we will discuss gapless surface states (SS) on sharp terminations of the spin-Hopf model. By a sharp termination we mean a Hamiltonian that is nearly identical to the corresponding Hamiltonian under periodic boundary conditions, except that all hopping matrix elements across the surface facet are turned off. 

In a $C_{nz}$ symmetric system with sharp terminations along the $z$-direction, each diagonal block $H_{l}(\boldsymbol{\Lambda})$ of the effective 1D tight-binding Hamiltonian at a $C_{nz}$ invariant $\boldsymbol{\Lambda}$ is a block Toeplitz matrix, i.e., the hopping matrix elements between unit cells at $\mathbf{R}$ and $\mathbf{R}^{\prime}$ depend on only  $\mathbf{R}-\mathbf{R}^{\prime}$ and satisfy $H_{l,\mathbf{R}\mathbf{R}^{\prime}}\equiv H_{l,\mathbf{R}-\mathbf{R}^{\prime}}=(H_{l,\mathbf{R}^{\prime}-\mathbf{R}})^{\dag}$, where $H_{l,\mathbf{R}-\mathbf{R}^{\prime}}$ is a matrix block. The spectral theorem of block Toeplitz matrices tells us that in the thermodynamic limit, the spectrum of $H_{l}(\boldsymbol{\Lambda})$ is bounded by the spectrum of its corresponding Bloch Hamiltonian $H_{l}(k_{z},\boldsymbol{\Lambda})$ \cite{nelson2021delicate}. 

For our spin-Hopf insulator model (which obeys the mutual disjoint condition) with sharp terminations, the above conclusion means that at each $C_{nz}$ invariant $\boldsymbol{\Lambda}$, any SS with angular momentum $l\in l_{v}$ ($l\in l_{c}$) must lie in the bulk valence (conduction) bands. Let us now focus on the top surface.  As shown in Fig.~\ref{fig:RTP} (a),  the bulk RTP  indicates that occupied (unoccupied) states at $M$ ($\Gamma$) will localize one layer higher along the $z$-direction than  the occupied (unoccupied) states at $\Gamma$ ($M$). Thus, on the top surface, i.e., the layer with $z=N$ as shown in Fig.~\ref{fig:RTP} (b), there will be an extra pair of occupied (unoccupied) states with angular momenta $+3/2, -3/2$ ($+1/2, -1/2$) at $M$ ($\Gamma$) due to the bulk RTP.
In each layer along the $z$-direction, there should always be four states in total having angular momenta $\pm 1/2, \pm 3/2$ at both $\Gamma$ and $M$ since the Hilbert space in each layer is spanned by the four basis orbitals \footnote{ As we can see from Fig.~\ref{fig:RTP} (b), this condition is satisfied at each bulk layer (i.e., layers with $z=2,3,\ldots, N-1$) }. Thus,  the extra states from the bulk RTP at $M$ ($\Gamma$) with angular momenta $\pm 3/2$ ($\pm 1/2$) should be compensated by a pair of SS with angular momenta $\pm 1/2\in l_{c}$ ($\pm 3/2 \in l_{v}$), as illustrated in Fig.~\ref{fig:RTP} (b). This means that the SS at $\Gamma$ ($M$) must lie in the bulk valence (conduction) bands in the energy spectrum, which guarantees the energy of SS must cross the bulk gap as one traverses from $\Gamma$ to $M$ as shown in Fig.~\ref{fig:RTP} (c). The same arguments can also be made for the bottom surface to imply the existence of SS on sharp terminations.

\subsection{Dirac dipole at topological phase transition point}
Finally, in this subsection we will consider the nature of the critical point separating a topological spin Hopf insulator and a trivial insulator. We will identify the topological phase transition point of the spin Hopf model to be a novel band degeneracy that we call a Dirac dipole in analogy with previous work on the conventional Hopf insulator.
Indeed, as discussed in Refs.~\onlinecite{alexandradinata2019actually,nelson2021delicate}, at the topological phase transition point between the trivial phase and the Hopf/RTP insulator in the MRW model [c.f. Eq.~\eqref{eq: MRW}], the low-energy $\mathbf{k}\cdot \mathbf{p}$ expansion around $\mathbf{k}=0$ yields a Berry dipole.
To understand why this is called a Berry dipole we can consider a sphere in momentum space that surrounds  $\mathbf{k}=0$  where  such a band degeneracy appears at the topological transition point $u=3$. Then, over the upper (lower) hemisphere the integral of the Berry curvature defined over the occupied subspace is quantized to be $2\pi$ ($-2\pi$). This has the structure of a dipole with opposite charges in opposite hemispheres. 

\begin{figure}[t!]
\centering
\includegraphics[width=1\columnwidth]{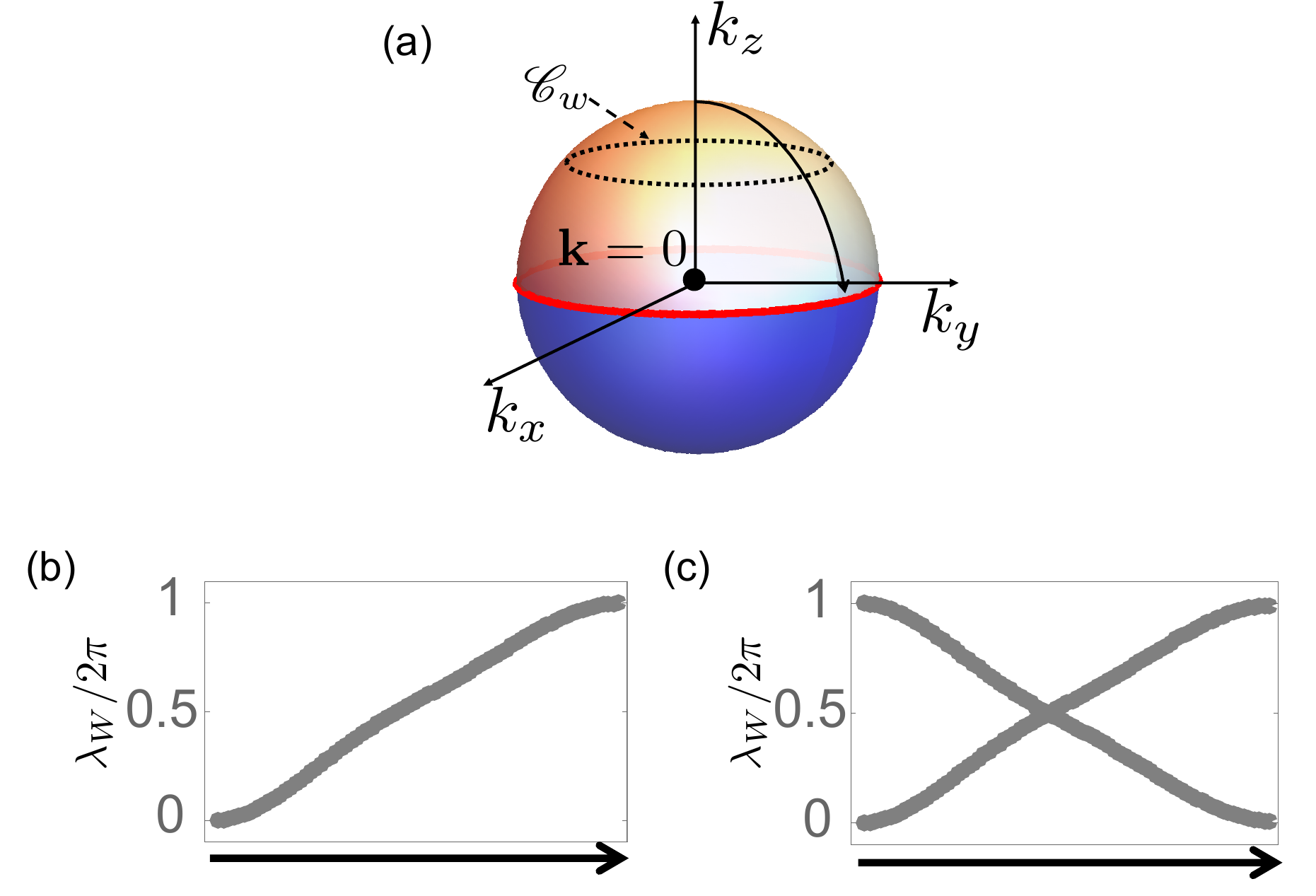}
\caption{(a) A sphere in momentum space surrounding the critical point at $\mathbf{k}=0$. Wilson loop spectra are calculated on the lattitude warps $\mathscr{C}_w.$ For MRW and spin-Hopf Hamiltonians with $u=3$, we plotted the spectra of Wilson loops along warps over the upper hemisphere versus the latitude (direction from north pole to equator is indicated by the black arrow) in (b) and (c). These spectra will be flipped when repeating the calculation from the equator to the south pole.}
\label{fig:BerryDiracdipole}
\end{figure}

For convenience one can explicitly correlate the Berry curvature calculations  with the winding of the Wilson loop spectrum over the hemisphere. For example,  as shown in Fig.~\ref{fig:BerryDiracdipole} (a), at each latitude on the upper hemisphere, we construct the Wilson loop operator along the warp $\mathscr{C}_{w}$, i.e., we take the path-ordered product of the projectors $p_{v}(\mathbf{k})$ onto the occupied subspace as shown in Eq.~\eqref{eq: WLO}.
In Fig.~\ref{fig:BerryDiracdipole} (b),  we plot the Wilson loop spectra of the MRW model (with $u=3$) versus the latitude from the north pole to the equator. We observe that the argument of the unit-modulus eigenvalue of $\hat{\mathcal{W}}_{w}$ changes by $+2\pi$, and thus  there is a positive winding of Wilson loop spectrum over the hemisphere, which implies a $+2\pi$-quantized Berry flux over the hemisphere. This winding flips sign if we repeat this process over the southern hemisphere moving from the south pole to the equator.  Note that since the MRW model is a two-band model with a single occupied band, each $\hat{\mathcal{W}}_{w}$ has only one unit-modulus eigenvalue. Since a Weyl point would have a $2\pi$ quantized Berry flux if we integrate over whole sphere surrounding it, the band touching of the Hopf insulator critical point is called a Berry dipole. Interestingly, the Berry dipole in the MRW model is protected by an emergent mirror symmetry along the $z$-direction \cite{nelson2021delicate,Sun2018conversion}. 

Since the spin Hopf model is a spin-doubled version of the MRW model, at the topological transition point $u=3$ there is a 4-fold band degeneracy at $\mathbf{k}=0$. For this degeneracy we expect to see a positive winding of the upper hemisphere Wilson loop spectrum in one spin sector, but a negative winding in the other spin sector [see Fig.~\ref{fig:BerryDiracdipole} (c)]. For the spin Hopf model, even though the net winding (i.e., the sum of the windings of the two unit-modulus eigenvalues) is zero, there is a nontrivial relative winding in the Wilson loop spectrum, i.e., the difference of the windings of the two unit-modulus eigenvalues is nonzero.  As a TR invariant analog of a Berry dipole, we call this novel band degeneracy a Dirac dipole because a single Dirac point degeneracy has a relative winding in the Wilson loop spectrum over the whole sphere surrounding it. In Appendix~\ref{app:symmetryconstraints}, we prove that this Dirac dipole, or equivalently, the relative winding in the Wilson loop spectrum over a hemisphere of the sphere surrounding the band degeneracy point \footnote{This means that we always assume the band degeneracy is preserved when discussing the stableness of the relative winding, i.e., we do not consider perturbations that lift the band degeneracy}, is
stabilized by TR, four-fold rotation, and emergent spatial inversion symmetries even when there are couplings between the two spin sectors.

\section{Conclusion and remark}
In conclusion, our work shows that the spin-Hopf insulator is a novel phase with a surface $\Z_{2}$ Kane Mele order. This topology has a higher-order character that is manifested by the helical hinge states in the presence of particle-hole inversion symmetry.
With four-fold rotation symmetry, the spin Hopf insulator is the first concrete model for delicate class-AII topological insulators that have nonzero RTP and associated gapless states on sharp boundaries. In addition, we have also explored the novel Dirac dipole band degeneracy at the phase transition point between a spin-Hopf insulator and a trivial insulator.

It is important to note that being non-magnetic (i.e., time-reversal invariant) makes the spin-Hopf insulators much more practical in different experimental platforms than Hopf insulators. On one hand, non-magnetic materials are more abundant in solid materials, which is the
reason why the realization of Chern insulator \cite{chang2013experimental} is much harder
than its time-reversal analog – the quantum spin Hall insulator \cite{konig2007quantum}.
On the other hand, in recent years, lots of highly-controllable metamaterials like photonic and acoustic crystals
has been developed to implement topological phases \cite{ozawa2019topological,xue2022topological}. Most of these systems naturally preserve time-reversal symmetry, and it usually requires highly nontrivial engineering to break the time-reversal symmetry \cite{wang2009observation,hafezi2013imaging,yang2015topological}. Meanwhile, non-magnetic systems are actively being studied for potential spintronic applications \cite{murakami2003dissipationless}, e.g., spin-charge conversion \cite{Zucchetti2018tuning} and spin Hall conductance \cite{guo2008intrinsic}. Therefore, we hope our works inspires further spintronic studies of non-magnetic topological insulators. 

Besides the phenomena of the spin-Hopf model discussed in this work, there are also some open questions for future research. (i) Though the topology of spin Hopf insulator with only TR symmetry looks like a boundary obstructed topological phase in the sense that the topology is protected by both the energy gap and the Wannier gap, the surface gap closing will not change the surface $\mathbbm{Z}_{2}$ that is unlike what in electric multipole insulators that have boundary obstructed topology \cite{benalcazar2017quantized,benalcazar2017electric,khalaf2021,zhu2021multi}. The relationship between the spin-Hopf insulators and boundary obstructed topological phases need further discussions. 
(ii)  We have mentioned that adding momentum-independent bands into the system while preserving the TR and particle-hole inversion symmetry will not destroy the surface $\Z_2$ invariant, and thus generalizes the spin-Hopf model to larger numbers of bands. However, it is unclear if there is a bulk homotopic invariant for larger band spin-Hopf insulators like that for the larger band Hopf insulator protected by particle-hole inversion symmetry discussed in Ref.~\onlinecite{Liu2017symmetry}. (iii) Recently, Ref.~\onlinecite{alexandradinata2022topological} has discussed the potential for a large transverse photovoltaic effect in class A delicate TIs with nonzero RTP. The photovoltaic effect (both transverse and longitudinal) and related phenomena in rotation symmetric class AII delicate TIs is still an open question.  We leave these questions for future research.

\section*{Acknowledgements}
P.Z, A.A, and T.L.H. thank Tomáš Bzdušek for insightful discussions. P.Z and T.L.H. thank the US Office of Naval Research (ONR) Multidisciplinary University Research Initiative (MURI) grant N00014-20- 1-2325 on Robust Photonic Materials with High-Order Topological Protection
for support.


\appendix
\begin{widetext}
\section{Adiabatic path connecting the spin Hopf insulator to a trivial band insulator without breaking the time-reversal symmetry}
\label{app:adiabaticpath}

Here, we explicitly construct the adiabatic path via three steps, and each of them can be described by a Bloch Hamiltonian with an adiabatic parameter.  The first step is
\begin{equation}
\begin{aligned}
&H_{1}(\mathbf{k}, t) =2(1-t)z_{1}z_{3} \mathbbm{1} \sigma_{x}+2 z_{2}z_{4} \tau_{z} \sigma_{x}
-2(1-t)z_{2}z_{3} \tau_{z} \sigma_{y}+2z_{1}z_{4}\mathbbm{1}\sigma_{y}
\\
&+\left(z_{1}^{2}+z_{2}^{2}-(1-t)^{2} z_{3}^{2}-z_{4}^{2}+t^{2} z_{3}^{2}\right) \mathbbm{1} \sigma_{z}
+2t(1-t) z_{3}^2\tau_{x} \sigma_{y}-2 t z_{4} z_{3} \tau_{x} \sigma_{x},
\end{aligned}
\end{equation}
where $t\in \left[0,1\right]$ and $H_{1}(\mathbf{k},0)=H_{sh}(\mathbf{k})$. The energy spectrum varies with $t$ as $E_1(\mathbf{k},t)=\pm\big(z_{1}^2+z_{2}^2+(1-t)^2z_{3}^2+z_{4}^2+t^2z_{3}^2\big)$, and we note that the bands with negative and positive energies are both two-fold degenerate. When $1<u<3$, $z_1$, $z_2$, $z_3$ and $z_{4}$ cannot be zero simultaneously. This is because if we want $z_1$, $z_2$ and $z_3$ to be zero, $k_{x}$, $k_{y}$ and $k_{z}$ must be $0$ or $\pi$, then only when $u=3,1,-3,-1$ can we have $z_{4}=0$. Thus, the bulk gap will never close for $t\in \left[0,1\right]$ if $1<u<3$. This step tunes the Bloch Hamiltonian so that at $t=1$, all the four by four matrices in the Bloch Hamiltonian $H_{1}(\bm{k},1)$ anti-commute with each other. 

The second step is 
\begin{equation}
\label{eq: 2ndstep}
\begin{aligned}
&H_{2}(\mathbf{k}, t)=\left(2(1-t) z_{1} z_{4}+t z_{2}\right) \mathbbm{1} \sigma_{y}+t z_{1} \tau_{y} \sigma_{x}
+\left(2(1-t)z_{2} z_{4}+tz_{3}\right) \tau_{z} \sigma_{x}-2 (1-t)z_{4} z_{3} \tau_{x} \sigma_{x}
\\
&+\left((1-t)^{2} z_{1}^{2}+(1-t)^2 z_{2}^{2}-z_{4}^{2}+(1-t)^2z_{3}^{2}\right) \mathbbm{1} \sigma_{z},
\end{aligned}
\end{equation}
where $t\in \left[0,1\right]$ and $H_{2}(\mathbf{k},0)=H_{1}(\mathbf{k},1)$. Since all five four by four matrices in the above Hamiltonian anti-commute with each other, the energy gap closes if and only if all five coefficients of the five matrices are simultaneously zero. Let us first consider $t=0,1$. At $t=0$, the energy spectrum is given by $E_{2}(\mathbf{k},0)=\pm(z_{1}^2+z_{2}^2+z_{3}^2+z_{4}^2)$, and hence the gap cannot close because $z_{i}$ for $i=1,2,3,4$ cannot be zero simultaneously. At $t=1$, $E_{2}(\mathbf{k},1)=\pm(z_{1}^2+z_{2}^2+z_{3}^2+z_{4}^4)$, so the gap cannot close for the same reason. Then, let us focus on $0<t<1$. If we want the second term and the fourth term to be zero, then we either have $z_{1}=z_{4}=0$ or $z_{1}=z_{3}=0$. If $z_{1}=z_{4}=0$, then the factor in the fifth term becomes $(1-t)^2 (z_{2}^2+z_{3}^2)$ which equals zero only when $z_{2}=z_{3}=0$. This is forbidden because $z_{i}$ for $i=1,2,3,4$ cannot be zero simultaneously when $1<u<3$. If $z_{1}=z_{3}=0$, then the factor in the first term becomes $t z_{2}$ which equals zero only when $z_{2}=0$. By substituting $z_{1}=z_{2}=z_{3}=0$ into the fifth term, the factor becomes $-z_{4}^2$ which equals zero only when $z_{4}=0$. Again, this cannot happen because $z_{i}$ for $i=1,2,3,4$ cannot be zero simultaneously when $1<u<3$.  Thus, we have proved that the energy gap remains open during this step.

Finally, the last step is \begin{equation}
\begin{aligned}
H_{3}(\mathbf{k},t)=(-z_{4}^2-tz_{3}^2-tz_{2}^2-tz_{1}^2)\mathbbm{1} \sigma_{z}+(1-t)z_{1}\tau_{y}\sigma_{x}+(1-t)z_{2} \mathbbm{1} \sigma_{y}+(1-t)z_{3} \tau_{z} \sigma_{x},
\end{aligned}
\end{equation}
where again $t\in \left[0,1\right]$ and $H_{3}(\mathbf{k},0)=H_{2}(\mathbf{k},1)$. For the same reason, the gap will never close in this step. As we can see, $H_{3}(\mathbf{k},1)=M\mathbbm{1}\sigma_{z}$, with mass $M=-\sum_{i=1}^{4}z_{i}^2<0$ for all $\mathbf{k}$ in the BZ. Thus, it is just a trivial band insulator. One can check explicitly that all these time-dependent Hamiltonians are symmetric under the TR symmetry $T=-i\tau_{y}\mathbbm{1}K$. 

In conclusion, this specific path indeed adiabatically connects the spin Hopf insulator with nontrivial surface $\mathbb{Z}_{2}$ number, to a trivial band insulator with trivial surface $\mathbb{Z}_{2}$ number without breaking the TR symmetry $T=-i\tau_{y}\mathbbm{1}K$. 

\section{Evolution of surface $\mathbbm{Z}_2$ invariant during our adiabatic path}
\label{app:evolution}
We numerically studied the evolution of the energy spectrum and the surface $\mathbbm{Z}_2$ invariant for insulators described by $H_1$, $H_2$, and $H_3$ with the $z$-direction open and the $x$, $y$-directions periodic. During the first step, i.e., $H_1$, the nontrivial $\mathbbm{Z}_2$ invariant on the top surface is in boundary-localized occupied bands, and the nontrivial $\mathbbm{Z}_2$ invariant on the bottom surface is in boundary-localized unoccupied bands bands as shown in Fig.~\ref{fig:surz2_h1}. At $H_1(t=1)=H_{2}(t=0)$, the surface gap closes on the top surface, and a phase transition occurs between the boundary-localized occupied bands and unoccupied bands on the top surface. Then, for $H_{2}(t)$ with $0<t<0.33$, the nontrivial $\mathbbm{Z}_2$ invariants are in boundary-localized unoccupied bands for both top and bottom surfaces. At $H_{2}(t=0.33)$ and $H_{2}(t=0.41)$, the surface closes along $\rm{X}-\Gamma-\rm{X}$ on the top and bottom surface subsequently. Again, phase transition happens at each of these two points, and we have the nontrivial $\mathbbm{Z}_2$ invariants in boundary-localized occupied bands for both top and bottom surfaces in $H_{2}(t)$ with $0.41<t<0.43$. Next, as discussed in the main text, the Bloch-Wannier bands touch with their neighbors at $H_{2}(t\approx 0.43)$, and pump a quantum of $\mathbbm{Z}_2$ invariant $\nu=1$ from one surface to another. Since this $\mathbbm{Z}_2$ pump only occurs in the occupied subspace in our adiabatic path, it leads to the 
removal of the nontrivial surface $\Z_{2}$ invariant.

\begin{figure}[h]
\centering
\includegraphics[width=0.8\columnwidth]{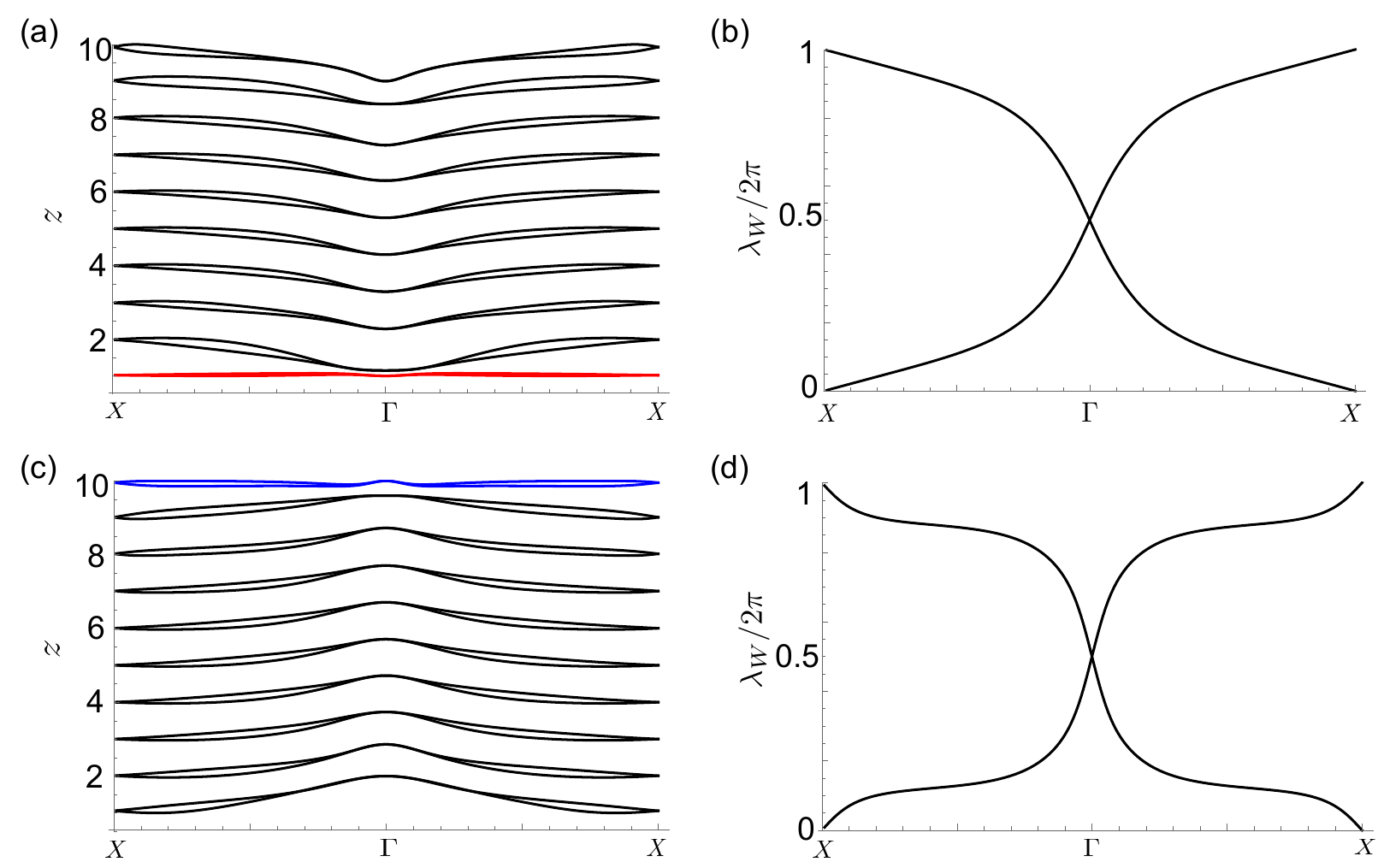}
\caption{(a) and (c) are Bloch-Wannier eigenbands of $Q\hat{z}Q$ and $P\hat{z}P$ for $H_{1}(t=0.5)$ with 10 unit cells along the $z$-direction, respectively. (b) shows the spectrum of Wilson loop along the $k_{y}$-direction (and plotted vs $k_x$) for Bloch-Wannier eigenbands localized at $z=1$ (marked as red in (a)) of $Q\hat{z}Q$; (d) shows the spectrum of Wilson loop along the $k_{y}$-direction (and plotted vs $k_x$) for Bloch-Wannier eigenbands localized at $z=10$ (marked as blue in (c)) of $P\hat{z}P$.}
\label{fig:surz2_h1}
\end{figure}

\section{Proof of Eq.~\eqref{eq:constraint_spinhopf} using the Wilson loop of the Bloch-Wannier bands}
\label{app: proofofeq31}
According to Eq.~\eqref{eq: WLO}, we can define and numerically calculate the Wilson loop operator for the Bloch-Wannier bands. We denote the Wilson loop operator along the $k_{y}$-direction (parallel transport in the $k_y$-direction, i.e., $\mathscr{C}$: $(k_{x},k_{y}=0)\rightarrow (k_{x},k_{y}=2\pi)$) of the occupied  Bloch-Wannier bands in the top half of the slab (derived by diagonalizing $P\hat{Z}P$) as $\hat{\mathcal{W}}_{v,t,y}(k_{x})$, such that
\begin{equation}
\hat{\mathcal{W}}_{v,t,y}(k_{x})=\prod_{k_{y}}^{2\pi \leftarrow 0}\sum_{z>0}\ket{w_{z}^{v}}\bra{w_{z}^{v}},
\end{equation}
where $z>0$ characterizes all Bloch-Wannier bands in the top half of the slab, and $\prod_{k_{y}}^{2\pi \leftarrow 0}$ denotes a path-ordered product along $k_{y}: 0\rightarrow 2\pi$.
Similarly, we denote the Wilson loop operator along the $k_{y}$-direction of the unoccupied  Bloch-Wannier bands in the bottom half of the slab (derived by diagonalizing $Q\hat{Z}Q$) as $\hat{\mathcal{W}}_{c,b,y}(k_{x})$, such that
\begin{equation}
\hat{\mathcal{W}}_{c,b,y}(k_{x})=\prod_{k_{y}}^{2\pi \leftarrow 0}\sum_{z<0}\ket{w_{z}^{c}}\bra{w_{z}^{c}},
\end{equation}
where $z<0$ characterizes all Bloch-Wannier bands in the bottom half of the slab. 

Then, the relation $\mathcal{C}^{\prime-1}\ket{w_{z}^{v}(k_{x},k_{y})}=\ket{w_{-z}^{c}(k_{x},k_{y})}$ discussed in the main text implies $\hat{\mathcal{W}}_{v,t,y}(k_{x})$ is unitarily equivalent to $\hat{\mathcal{W}}_{c,b,y}^{\star}(k_{x})$, where $^{\star}$ is the complex conjugate due to the complex conjugate operator $K$ in the particle-hole inversion operator. As mentioned in the main text, it is known that the $\Z_{2}$ Kane-Mele invariant can be derived by counting the parity of the number of intersections between a line parallel to the $k_{x}$-axis and the spectrum of the Wilson loop $\lambda_{W}(k_{x})$  \cite{PhysRevB.74.195312,bernevig2013topological,asboth2016short}. Let 
the spectrum of $\hat{\mathcal{W}}_{v,t,y}(k_{x})$, $\lambda_{W_{vt}}(k_{x})$, cross an arbitrary line parallel to the $k_{x}$-axis in the region $\Gamma$-$X$ an odd number of times and thus have $\nu_{v}(\hat{z})=1 \ \rm{mod}\ 2$. Then, given the unitary equivalence between $\hat{\mathcal{W}}_{v,t,y}(k_{x})$ and $\hat{\mathcal{W}}^{\star}_{c,b,y}(k_{x})$,  the spectrum of $\hat{\mathcal{W}}_{c,b,y}(k_{x})$, $\lambda_{W_{cb}}(k_{x})=\lambda^{\star}_{W_{vt}}(k_{x})$, must also cross such a line an odd number of times and thus have $\nu_{c}(-\hat{z})=1 \ \rm{mod}\ 2=\nu_{v}(+\hat{z})$. By a similar procedure, we can also get $\nu_{v}(-\hat{z})=\nu_{c}(+\hat{z})$.

\section{symmetry-protected Dirac dipole}
\label{app:symmetryconstraints}
To discuss the Dirac dipole protected by TR, four-fold rotation, and emergent spatial inversion symmetry, we first write down the $\mathbf{k}\cdot\mathbf{p}$ expansion of the spin Hopf model at transition point $u=3$ around the $\mathbf{k}=0$ point:
\begin{equation}
\label{eq:Diracdipole}
\begin{aligned}
H_{\Gamma}(\mathbf{k})=2 k_{x} k_{z} \mathbbm{1} \sigma_{x}-2 k_{y} k_{z} \tau_{z} \sigma_{y}+\left(k_{x}^{2}+k_{y}^{2}-k_{z}^{2}\right) \mathbbm{1} \sigma_{z},
\end{aligned}
\end{equation}
which has an emergent inversion symmetry $\mathcal{I}H_{\Gamma}(\mathbf{k})\mathcal{I}^{-1}=H_{\Gamma}(-\mathbf{k})$, where $\mathcal{I}=\mathbbm{1}_{4\times 4}$.  

To show that the Dirac dipole in the spin-Hopf model is robust, we first  analyze the symmetry constraints on the Wilson loop operator along a warp on the sphere surrounding $\mathbf{k}=0$ point [see Fig.~\ref{fig:BerryDiracdipole} (a) in the main text]. Without loss of generality, let us consider a warp at a given $k_{z}$, and set the starting point for the Wilson loop operator to be $\mathbf{k}_{0}$. Then, the Wilson loop operator is
\begin{equation}
\hat{\mathcal{W}}_{w}=P(\mathbf{k_{0}})\ldots P(\mathbf{k_{0}}+2\delta \mathbf{k})P(\mathbf{k_{0}}+\delta \mathbf{k})P(\mathbf{k_{0}}).
\end{equation}
A time reversal ($T=\tau K$, $T^2=-1$) times spatial inversion symmetry ($\mathcal{I}$) acts on this Wilson loop operator as
\begin{equation}
 T\mathcal{I} \hat{\mathcal{W}}_{w} ( T\mathcal{I})^{-1}= \tau\mathcal{I}^{\star} \hat{\mathcal{W}}^{\star}_{w} ( \tau\mathcal{I}^{\star})^{-1}=\hat{\mathcal{W}}_{w},
\end{equation}
because $ T\mathcal{I} P(\mathbf{k}) (T\mathcal{I})^{-1}=P(\mathbf{k})$. This means that $\hat{\mathcal{W}}^{\star}_{w}$ and $\hat{\mathcal{W}}_{w}$ are similar to each other, and thus they have the same spectrum. Then, for each eigenvalue of $\hat{\mathcal{W}}_{w}$, its complex conjugate must also be an eigenvalue of $\hat{\mathcal{W}}_{w}$. Therefore, the unit-modulus eigenvalues of $\hat{\mathcal{W}}_{w}$ can only be $\pm 1$ or appear as a pair of complex conjugate numbers. This constraint guarantees that the spectrum of Wilson loop along a warp is always symmetric with respect to $\lambda_{w}=0,\pi$, and thus the degeneracy in the spectrum can only happen at $\lambda_{w}=0,\pi$ . If we furthermore show that the degeneracy at $\lambda_{w}=0,\pi$ in Fig.~\ref{fig:BerryDiracdipole} (c) cannot be lifted through hybridizing, then we can prove the relative winding is robust. 

Next, we show that the four-fold rotation symmetry can protect the degeneracy at $\lambda_{w}=0,\pi$. From the four-fold rotation symmetry $C_{4z}$, we can construct an operator $O$ such that $O^{4}=-\hat{\mathcal{W}}_{w}^{\dag}$ and thus $[O,\hat{\mathcal{W}}^{\dag}_{w}]=0$. Explicitly, the operator $O$ takes the form
\begin{equation}
\label{eq:O}
O(\mathbf{k}_{0})=\left(\prod_{\mathbf{k}=R_{4}\mathbf{k}_{0}}^{ \mathbf{k}_{0}}P(\mathbf{k}^{\prime})\right)C_{4z},
\end{equation}
where $\mathbf{k}_{0}$ is the starting point of the Wilson loop operator on a given warp, and $R_{4}$ rotates the momentum by $\pi/2$, and $O^{4}=-\hat{\mathcal{W}}_{w}^{\dag}$ can be proved through the basic relationship $C_{4z}P(\mathbf{k})C_{4z}=P(R_{4}\mathbf{k})$:
\begin{equation}
\begin{aligned}
O^{4}(\mathbf{k_{0}})&=\left(\prod_{\mathbf{k}=R_{4}\mathbf{k}_{0}}^{ \mathbf{k}_{0}}P(\mathbf{k}^{\prime})\right)C_{4z}\left(\prod_{\mathbf{k}=R_{4}\mathbf{k}_{0}}^{ \mathbf{k}_{0}}P(\mathbf{k}^{\prime})\right)C_{4z}\left(\prod_{\mathbf{k}=R_{4}\mathbf{k}_{0}}^{ \mathbf{k}_{0}}P(\mathbf{k}^{\prime})\right)C_{4z}\left(\prod_{\mathbf{k}=R_{4}\mathbf{k}_{0}}^{ \mathbf{k}_{0}}P(\mathbf{k}^{\prime})\right)C_{4z}
\\
&=\left(\prod_{\mathbf{k}=R_{4}\mathbf{k}_{0}}^{ \mathbf{k}_{0}}P(\mathbf{k}^{\prime})\right)\left(\prod_{\mathbf{k}=R^2_{4}\mathbf{k}_{0}}^{ R_{4}\mathbf{k}_{0}}P(\mathbf{k}^{\prime})\right)C^2_{4z}\left(\prod_{\mathbf{k}=R_{4}\mathbf{k}_{0}}^{ \mathbf{k}_{0}}P(\mathbf{k}^{\prime})\right)C_{4z}\left(\prod_{\mathbf{k}=R_{4}\mathbf{k}_{0}}^{ \mathbf{k}_{0}}P(\mathbf{k}^{\prime})\right)C_{4z}
\\
&=\left(\prod_{\mathbf{k}=R_{4}\mathbf{k}_{0}}^{ \mathbf{k}_{0}}P(\mathbf{k}^{\prime})\right)\left(\prod_{\mathbf{k}=R^2_{4}\mathbf{k}_{0}}^{ R_{4}\mathbf{k}_{0}}P(\mathbf{k}^{\prime})\right)\left(\prod_{\mathbf{k}=R^{3}_{4}\mathbf{k}_{0}}^{ R^{2}_{4}\mathbf{k}_{0}}P(\mathbf{k}^{\prime})\right)C^{3}_{4z}\left(\prod_{\mathbf{k}=R_{4}\mathbf{k}_{0}}^{ \mathbf{k}_{0}}P(\mathbf{k}^{\prime})\right)C_{4z}
\\
&=\left(\prod_{\mathbf{k}=R_{4}\mathbf{k}_{0}}^{ \mathbf{k}_{0}}P(\mathbf{k}^{\prime})\right)\left(\prod_{\mathbf{k}=R^2_{4}\mathbf{k}_{0}}^{ R_{4}\mathbf{k}_{0}}P(\mathbf{k}^{\prime})\right)\left(\prod_{\mathbf{k}=R^{3}_{4}\mathbf{k}_{0}}^{ R^{2}_{4}\mathbf{k}_{0}}P(\mathbf{k}^{\prime})\right)\left(\prod_{\mathbf{k}=\mathbf{k}_{0}}^{ R_{4}^{3}\mathbf{k}_{0}}P(\mathbf{k}^{\prime})\right)C^{4}_{4z}=-\hat{\mathcal{W}}^{\dag}_{w},
\end{aligned}
\end{equation}
where the factor $-1$ is because of $C^{4}_{4z}=-1$ in our case. It is straightforward to see that $O^{4}=-\hat{\mathcal{W}}_{w}^{\dag}$ directly indicates  $[O,\hat{\mathcal{W}}_{w}]=0$. Thus, if two degenerate eigenstates of $\hat{\mathcal{W}}^{\dag}_{w}$ (and thus of $\hat{\mathcal{W}}_{w}$) have different eigenvalues of $O$, their degeneracy cannot be lifted by any $C_{4z}$ symmetry-preserving perturbations  \cite{Alex2018semiclassical}. 

With these symmetry constraints, we are now ready to show that in the spin-Hopf model, the relative winding in the spetrum of Wilson loop over a hemisphere is robust as long as the symmetries and the band degeneracy are preserved.
First, at the north pole of the sphere, the Hamiltonian simplifies and the Wilson loop operator at the fixed momentum $(0,0,k_{z0})$ has  two unit-modulus eigenvalues that are degenerate and equal to $1$ (i.e., $\lambda_{W}=0$).

Second, for the crossing points at $\lambda_{W}=\pi$ [see Fig.~\ref{fig:BerryDiracdipole} (c)], because $O^{4}=-\hat{\mathcal{W}}^{\dag}_{w}$ the eigenvalues of $O$ can  take only discrete values $\exp(i n\pi/2)$ with $n=0,1,2,3$. For the spin Hopf model, the two eigenstates of $\hat{\mathcal{W}}_{w}$ with unit-modulus eigenvalues are eigenstates of $O$ with eigenvalues $\exp(i \pi/2)$ and $\exp(i 3\pi/2)$. Continuous deformations preserving symmetries and the band degeneracy can never change the eigenvalues of $O$ given that they only take discrete values. Since $\exp(i \pi/2)\neq\exp(i 3\pi/2)$, this crossing at $\lambda_{W}=\pi$ is then robust against any symmetry preserving perturbations. 

Finally, for $\hat{\mathcal{W}}_{w}$ along the equator of the spin Hopf model with $u=3$, the two unit-modulus eigenvalues are both $1$ (i.e., $\lambda_{W}=0$). These two eigenstates of $\hat{\mathcal{W}}_{w}$ are eigenstates of $O$ with eigenvalues $\exp(i \pi/4)$ and $\exp(i 7\pi/4)$. Following the same argument in the last paragraph, the degeneracy at $\lambda_{W}=0$ in the spectrum of $\hat{\mathcal{W}}_{w}$ along the equator is also robust against any symmetry preserved perturbations. Thus, starting from the spin Hopf model with spin-$U(1)$ symmetry, the relative winding in the spectrum of the Wilson loops is robust against any perturbations that preserve the symmetries.

We note that by assumption we consider only band degeneracy preserving perturbations when discussing the Dirac dipole. However, symmetry preserved perturbation terms do not automatically preserve the band degeneracy. For clarification,
let us now show all possible perturbation terms for the 4-band spin-Hopf model preserving both the band degeneracy and the symmetries up to second order in $k_{x}, k_{y}, k_{z}$, which are perturbation terms we focused on.

The spatial inversion symmetry precludes all terms linear in $k_{x}, k_{y}, k_{z}$. Under TR transformation, five out of fifteen $4\times 4$ Hermitian matrices are invariant, and all others gain an extra minus sign. The five TR invariant matrices are $\tau_{x}\sigma_{y}$, $\tau_{y}\sigma_{y}$, $\tau_{z}\sigma_{y}$, $\mathbbm{1}\sigma_{x}$, and $\mathbbm{1}\sigma_{z}$. Under $C_{4z}=\exp\left(i\frac{\pi}{2}\tau_{z}(\frac{\sigma_{z}}{2}+\mathbbm{1})\right)$, the transformations of the fifteen $4\times 4$ Hermitian matrices are summarized in Table~\ref{tab:transform}. Then, all possible terms that preserve the TR, four-fold rotation, spatial inversion symmetries and the band degeneracy at $\mathbf{k}=0$ are:

\begin{equation}
\label{eq: pert}
[\alpha_1 k_{x}k_{y}+\beta_1(k_{x}^2-k_{y}^2)]\tau_{x}\sigma_{y},\quad [\alpha_2 k_{x}k_{y}+\beta_{2}(k^2_{x}-k^2_{y})]\tau_{y}\sigma_{y}, \quad
k_{x}k_{z}\tau_{z}\sigma_{y}+k_{y}k_{z}\mathbbm{1}\sigma_{x},\quad
k_{x}k_{z}\mathbbm{1}\sigma_{x}-k_{y}k_{z}\tau_{z}\sigma_{y}, \quad
[a(k_{x}^2+k_{y}^2)+bk_{z}^2] \mathbbm{1}\sigma_{z},
\end{equation}
where only the first two terms couple different spin sectors. We have numerically confirmed that the relative winding is indeed robust under these perturbation terms (especially the first two terms that break the spin-$U(1)$ symmetry). Thus, even if we move away from the simple spin-$U(1)$ symmetric case, we can still have the Dirac dipole band degeneracy.

\begin{table}[h]
\caption{Transformation of fifteen $4\times 4$ Hermitian matrices under $C_{4z}$}
\label{tab:transform}
\renewcommand{\arraystretch}{1.5}
\begin{tabular}{p{0.1\textwidth}|p{0.19\columnwidth}p{0.19\columnwidth}p{0.19\columnwidth}p{0.19\columnwidth}}
\hline\hline
$M$
      & $\tau_{x}\sigma_{x}$  & $\tau_{x}\sigma_{y}$ & $\tau_{x}\sigma_{z}$
      & $\tau_{x}\mathbbm{1}$\\ 
\hline
$C_{4z}MC_{4z}^{-1}$ & $-\tau_{x}\sigma_{x}$ &  $-\tau_{x}\sigma_{y}$ & $\tau_{y}\mathbbm{1}$ & $\tau_{y}\sigma_{z}$\\ 
\hline\hline
$M$
      & $\tau_{y}\sigma_{x}$  & $\tau_{y}\sigma_{y}$ & $\tau_{y}\sigma_{z}$
      & $\tau_{y}\mathbbm{1}$\\ 
\hline
$C_{4z}MC_{4z}^{-1}$ & $-\tau_{y}\sigma_{x}$ &  $-\tau_{y}\sigma_{y}$ & $-\tau_{x}\mathbbm{1}$ & $-\tau_{x}\sigma_{z}$\\ 
\hline\hline
$M$
 & $\tau_{z}\sigma_{x}$  & $\tau_{z}\sigma_{y}$ & $\tau_{z}\sigma_{z}$
      & $\tau_{z}\mathbbm{1}$\\ 
\hline
$C_{4z}MC_{4z}^{-1}$& $-\mathbbm{1}\sigma_{y}$ &  $\mathbbm{1}\sigma_{x}$ & $\tau_{z}\sigma_{z}$ & $\tau_{z}\mathbbm{1}$\\ 
\hline\hline
$M$
 & $\mathbbm{1}\sigma_{x}$  & $\mathbbm{1}\sigma_{y}$ & $\mathbbm{1}\sigma_{z}$
      & \\ 
\hline
$C_{4z}MC_{4z}^{-1}$ & $-\tau_{z}\sigma_{y}$ &  $\tau_{z}\sigma_{x}$ & $\mathbbm{1}\sigma_{z}$& \\ 
\hline\hline
\end{tabular}
\end{table}

\end{widetext}

\bibliography{apssamp}



\end{document}